\documentclass[lettersize,journal]{IEEEtran}
\usepackage{amsmath,amsfonts}
\usepackage{algorithmic}
\usepackage{algorithm}
\usepackage{array}
\usepackage[caption=false,font=normalsize,labelfont=sf,textfont=sf]{subfig}
\usepackage{textcomp}
\usepackage{stfloats}
\usepackage{url}
\usepackage{verbatim}
\usepackage{graphicx}
\usepackage{cite}
\begin{document}

\title{Unlocking Directional Radiation in Pinching-Antenna Systems: Geometry-Aware Design and Experimental Verification}

\author{Haoyang Li, Weidong Liu, \IEEEmembership{Graduate Student Member,~IEEE}, Zhongliang Li, \IEEEmembership{Student Member,~IEEE}, \\Gaojie Chen, \IEEEmembership{Senior Member,~IEEE}, Zheng Yang, \IEEEmembership{Senior Member,~IEEE} and Zhiguo Ding, \IEEEmembership{Fellow,~IEEE}\vspace{-2em}
        %<-this % stops a space

% \thanks{This work was supported in part by the Fundamental and Interdisciplinary Disciplines Breakthrough Plan of the Ministry of Education of China under Grant JYB2025XDXM406, and by the Fundamental Research Funds for the Central Universities, Sun Yat-sen University under Grant No.24hytd010.}
\thanks{Haoyang Li, Zhongliang Li and Gaojie Chen are with the School of Flexible Electronics (SoFE), Sun Yat-sen University, Shenzhen, Guangdong 518107, China (e-mail: hylifj@foxmail.com; lizhliang6@mail2.sysu.edu.cn; chengj235@mail.sysu.edu.cn).
}
\thanks{Weidong Liu is with the School of Telecommunications Engineering, Xidian University, Xi'an 710071, China, and also with the School of Flexible Electronics (SoFE), Sun Yat-sen University, Shenzhen, Guangdong 518107, China (e-mail: wdlui@stu.xidian.edu.cn).
}
\thanks{Zheng Yang is with the Fujian Provincial Engineering Technology Research Center of Photoelectric Sensing Application, Fujian Normal University, Fuzhou 350117, China, (e-mail: zyfjnu@163.com).
}
\thanks{Zhiguo Ding is with the School of Electrical and Electronic Engineering, Nanyang Technological University, Singapore 639798 (e-mail:zhiguo.ding@ntu.edu.sg).
}
}

% The paper headers
% \markboth{Journal of \LaTeX\ Class Files,~Vol.~14, No.~8, August~2021}%
% {Shell \MakeLowercase{\textit{et al.}}: A Sample Article Using IEEEtran.cls for IEEE Journals}

% \IEEEpubid{0000--0000/00\$00.00~\copyright~2021 IEEE}
% Remember, if you use this you must call \IEEEpubidadjcol in the second
% column for its text to clear the IEEEpubid mark.

\maketitle

\begin{abstract}
Pinching-antenna systems (PASS) have recently attracted growing interest as a flexible architecture for creating ``last-meter" line-of-sight wireless links through dielectric waveguides and reconfigurable radiation points. While modeling pinching antennas (PAs) as isotropic point radiators has enabled tractable analyses and demonstrated the performance gains of PASS, their practical radiation characteristics remain underexplored. This article investigates PASS from the perspective of PA geometry. Starting from the physical coupling principle, we explain why PA shape affects the induced polarization current and incorporate directional gain into the channel model. The full-wave simulations are conducted to show how different PA geometries and orientations reshape the internal field distribution and far-field radiation pattern. A 60 GHz prototype video transmission experiment is further presented to demonstrate the link-level impact of changing PA states. Finally, promising applications enabled by geometry-aware directional PASS are highlighted.
\end{abstract}

% \begin{IEEEkeywords}
% Article submission, IEEE, IEEEtran, journal, \LaTeX, paper, template, typesetting.
% \end{IEEEkeywords}

\section{Introduction}
Future 6G networks are moving toward millimeter-wave and terahertz bands to support higher data rates, denser devices, and richer sensing services. Yet these bands also intensify two long-standing wireless challenges: severe free-space path loss and line-of-sight (LoS) blockage. Existing technologies can help, but only to a limited extent. Massive multiple-input multiple-output (MIMO) can provide beamforming gain but remains tied to fixed antenna locations, reconfigurable intelligent surfaces (RISs) may suffer from cascaded path loss, and fluid or movable antennas usually reconfigure positions within a wavelength-scale range \cite{ding2025flexible}. This raises the question of whether the network can flexibly bring radiation closer to the serving user rather than only shaping it from afar.

Pinching-antenna systems (PASS) offer such a possibility. A PASS uses a dielectric waveguide as a low-loss guided path that can also be locally used as antennas by activating small dielectric elements, referred to as pinching antennas (PAs), as flexible radiation points along the waveguide \cite{fukuda2022pinching}. In this way, the waveguide carries the signal close to the terminal, and the PA creates a strong ``last-meter" LoS link from nearby infrastructure such as an indoor ceiling, a corridor wall, or a factory rail. In this sense, PASS is not a conventional RF-fed antenna array, but a guided--wireless architecture that carries energy along a waveguide and radiates it where needed. By shortening the free-space distance, avoiding local blockers, and activating low-cost radiation points on demand, PASS provides a flexible and scalable way to expand high-frequency coverage without requiring dense access-point deployments.

By assuming that each activated PA is treated as an isotropic point radiator, an extensive number of existing PASS studies have been carried out to demonstrate the superior performance of PASS over conventional communication systems. This isotropic abstraction is useful for system-level analysis and optimization, but it also conceals an important hardware fact \cite{xu2025rate,wang2025antenna}. The early PASS prototype in \cite{fukuda2022pinching} demonstrated 60 GHz video transmission and verified the feasibility of PASS, but the radiation characteristics of individual PA elements were not fully investigated. In practice, a PA is a dielectric structure whose radiation may exhibit directionality, with its pattern depending on the PA geometry and orientation. This geometry-dependent radiation introduces an additional design degree of freedom, enabling PASS to regulate not only the radiation location but also the spatial direction in which the extracted guided energy is delivered. In multi-user PASS scenarios, such directionality can be further exploited to enhance communication performance, for example, by mitigating co-channel interference and facilitating spatial separation among users.

This observation suggests two complementary geometry-aware design strategies for PASS. The first is to engineer PAs with near-isotropic radiation characteristics, such that practical PA hardware better matches the isotropic-radiator assumption underlying existing location-tuning and beamforming methods. The second is to intentionally design directional PAs whose geometries and orientations concentrate radiated energy toward desired service regions, thereby introducing element-level pattern control beyond PA placement.

These two strategies motivate a geometry-aware view of PASS. Existing location-tuning algorithms determine where a PA radiates along the waveguide, while the resulting radiation pattern is also governed by the geometry and orientation of the attached PA element. Therefore, PA geometry and orientation should be considered controllable design dimensions alongside PA positioning. In other words, a PA should not only be placed at the right point, but also shaped and oriented for the desired radiation direction. Recent geometry-dependent and directional PA models also highlight the need for physics-based PASS channel modeling \cite{li2026geometry,zhang2025directional,zubair2026closed}.

In this article, we first connect the evanescent-field coupling mechanism with a geometry-dependent channel model, clarifying how directional gain enters the channel coefficient. Full-wave simulation results are then presented to compare representative PA shapes and show how PA rotation and orientation affect the radiation direction. A 60 GHz prototype video transmission experiment is further presented to verify the resulting signal variations. Finally, we discuss promising application scenarios where geometry-aware directional PAs can offer benefits beyond location tuning.

\section{Fundamentals of Pinching Antennas}
To move beyond the idealized isotropic-radiator assumption, this section connects PA radiation behavior with its physical structure. This section first explains the radiation mechanism of PAs, formulates a geometry-aware channel model that accounts for directional gain, and then illustrates how PA geometry affects radiation through representative PA shapes.

\subsection{Physical Principle}
In PASS, a dielectric waveguide carries the RF signal, while small dielectric elements, i.e., PAs, locally extract guided energy and radiate it into free space. This process follows the standard evanescent-field coupling principle, in which the guided mode is mainly confined in the waveguide while its evanescent field extends outside the boundary and transfers energy to a nearby dielectric element \cite{marcuse1974theory}. Introducing a PA into this region creates a local radiation point, whereas removing it suppresses local radiation and restores guided transmission along the waveguide. This reversible coupling enables flexible PA activation along the waveguide.

The following discussion turns to the radiation process after guided energy is coupled into the PA. According to the volume equivalence principle \cite{balanis2012advanced}, the far-field radiation is determined by the equivalent polarization current $\mathbf{J}_{\mathrm{eq}} = j\omega(\epsilon_{\mathrm{PA}} - \epsilon_0)\mathbf{E}_{\mathrm{inc}}$ induced within the dielectric volume, where $\omega$ is the angular frequency, $\epsilon_{\mathrm{PA}}$ and $\epsilon_0$ denote the PA and vacuum permittivities, respectively, and $\mathbf{E}_{\mathrm{inc}}$ is the incident electric field coupled from the waveguide into the PA. Since this current distribution depends on the PA boundary and internal field propagation, changing the PA geometry can reshape the far-field pattern \cite{li2026geometry}. The resulting directionality can therefore be reconfigured by changing the PA shape and further incorporated into the PA-user channel model.

\begin{figure} [t!]
        \centering
        \includegraphics[width= 3.24in]{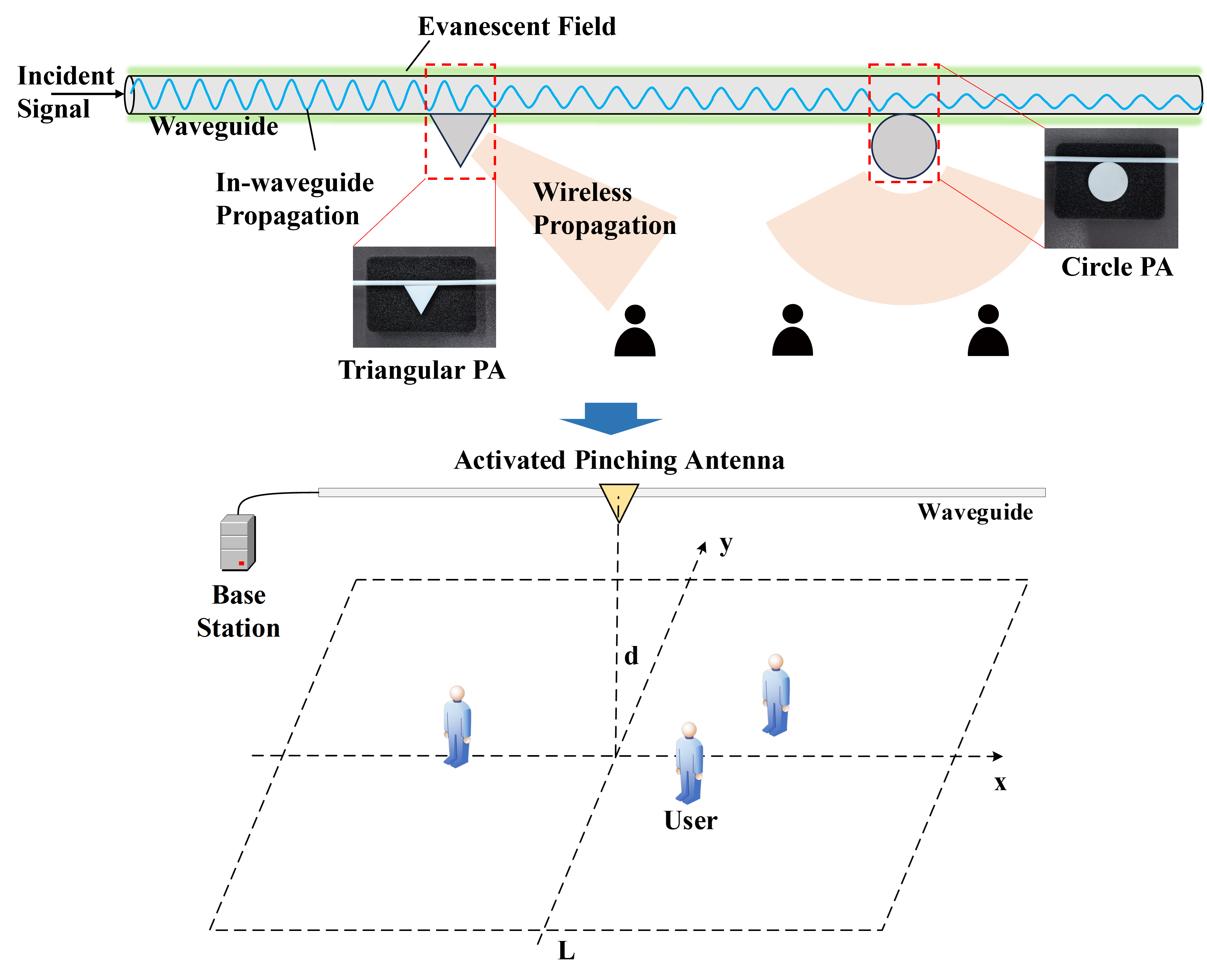}
        \caption{Illustration of PASS with different geometries of PAs.}\label{illustration.Fig}
\end{figure}

\subsection{Geometry-Dependent Channel Gain}
As illustrated in Fig.~\ref{illustration.Fig}, a dielectric waveguide deployed at height $d$ can accommodate multiple PAs with different geometries to serve users within a square coverage area of side length $L$. For analytical clarity, we focus on the channel between a single PA and one user. The composite channel from the base station to the user comprises two cascaded stages: in-waveguide propagation and free-space propagation \cite{ding2025flexible}. For analytical tractability, many existing works model PAs as isotropic radiators, i.e., the radiation is assumed uniform in all directions. To further capture practical PA behavior, the PA geometry can be taken into account because it governs the spatial distribution of the induced polarization current, which produces a direction-dependent radiation pattern. To incorporate this directional gain into the channel model, the channel coefficient can be expressed as \cite{li2026geometry}, \cite{zhang2025directional}
\setlength\abovedisplayskip{3pt} 
\setlength\belowdisplayskip{3pt}
\begin{equation}\label{pinch_channel.eq}
  \begin{aligned}
    h = \frac{\eta \kappa}{D} \sqrt{G_\mathcal{G} (\theta, \phi)} e^{-j(\frac{2\pi}{\lambda} D +\frac{2\pi}{\lambda_g} L_w)},
  \end{aligned}
\end{equation}
where $\lambda$ and $\lambda_g$ denote the free-space and guided wavelengths, respectively, $\eta$ is the wavelength-dependent amplitude factor, and $D$ is the PA-to-user distance. The coupling factor $\kappa$ quantifies the fraction of guided-wave power extracted by the PA, and $L_w$ is the feed-to-PA distance along the waveguide. $\theta$ and $\phi$ denote the elevation and azimuth angles toward the user, respectively, with the radiating point taken as the origin. The term $G_{\mathcal{G}}(\theta,\phi)$ denotes the directional gain of a PA toward the user and can be obtained from full-wave simulations for arbitrary PA configurations, where $\mathcal{G}$ specifies the PA geometry, orientation, and rotation state. From a communication perspective, $D$ characterizes the distance-dependent path loss, whereas $(\theta,\phi)$ characterize the angular state of the PA-user link. Consequently, two users at similar distances from the PA may still experience different channel gains if they are located in different directions of the same PA.

The key message of \eqref{pinch_channel.eq} is that the PA configuration $\mathcal{G}$, including its shape, rotation state, and installation orientation, enters the communication channel through the directional gain $G_{\mathcal{G}}(\theta,\phi)$. Thus, even with the PA-user distance and coupling factor fixed, changing $\mathcal{G}$ can alter the channel gain by reshaping the radiation pattern observed in the user direction. Physically, the PA geometry determines the propagation paths of the coupled wave inside the dielectric volume. Differences in path length then produce a non-uniform phase distribution across the radiating aperture, which governs the far-field beam direction \cite{li2026geometry}. Conventional PA location tuning mainly adjusts the PA-user distance and propagation phase. Consequently, a user located in a low-gain direction may still experience weak reception, even when the PA position is favorable. This observation yields two complementary design strategies, as conceptually depicted in Fig.~\ref{illustration.Fig}: first, engineering PA geometries to deliver near-isotropic radiation, so as to better match the isotropic assumption adopted by existing location-tuning-based beamforming algorithms \cite{li2025total}; second, deliberately tailoring PA geometry (e.g., triangular) to steer radiation toward intended coverage areas, thereby introducing a hardware-level degree of freedom for PA systems independent of PA location tuning.

\subsection{Effect of PA Shape}

\begin{figure*} [t!]
        \centering
        % Row 1: E-field distributions
        \subfloat[]{\makebox[0.24\textwidth][c]{\includegraphics[height=1.30in]{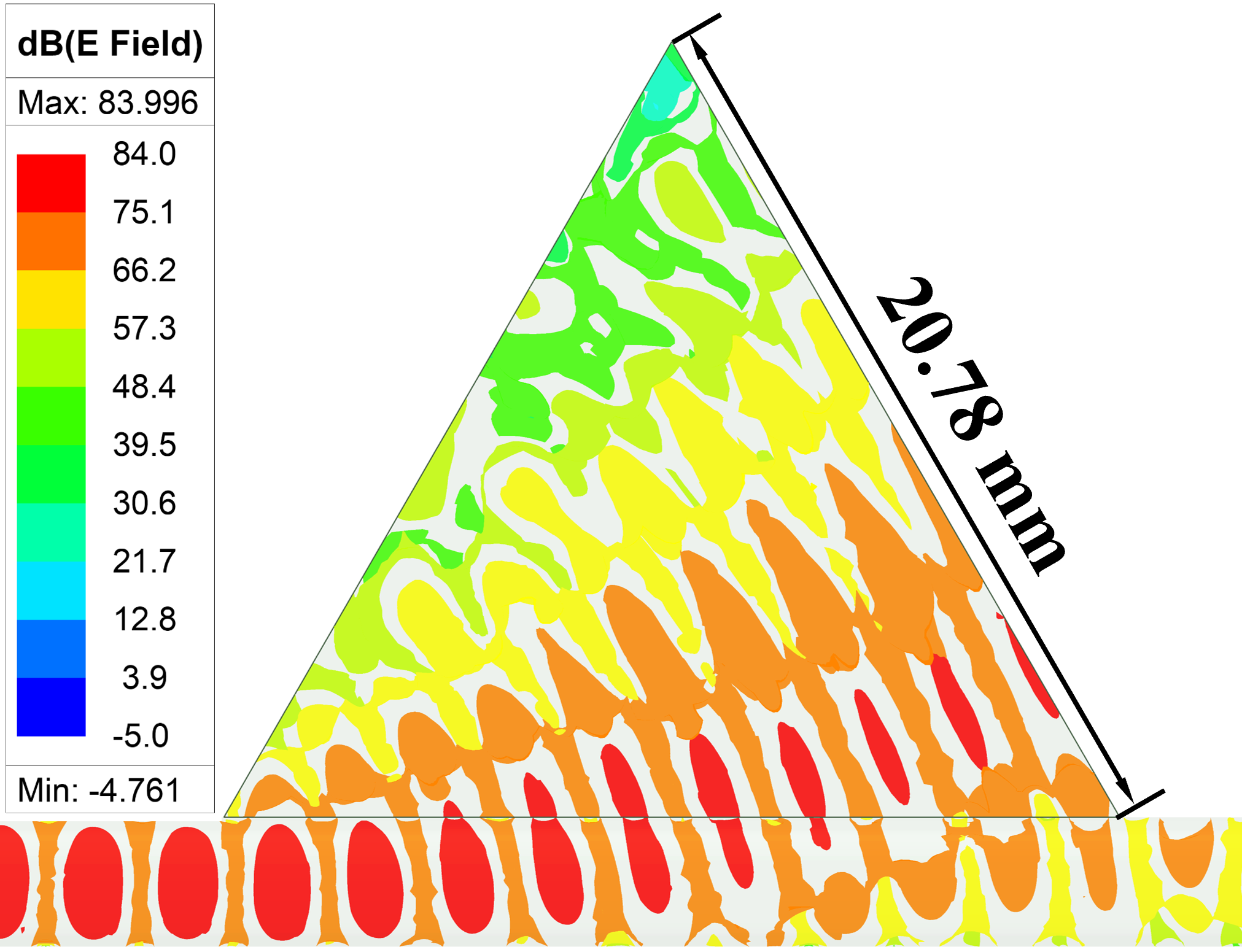}}\label{tri.Fig}}
        \hfil
        \subfloat[]{\makebox[0.24\textwidth][c]{\includegraphics[height=1.30in]{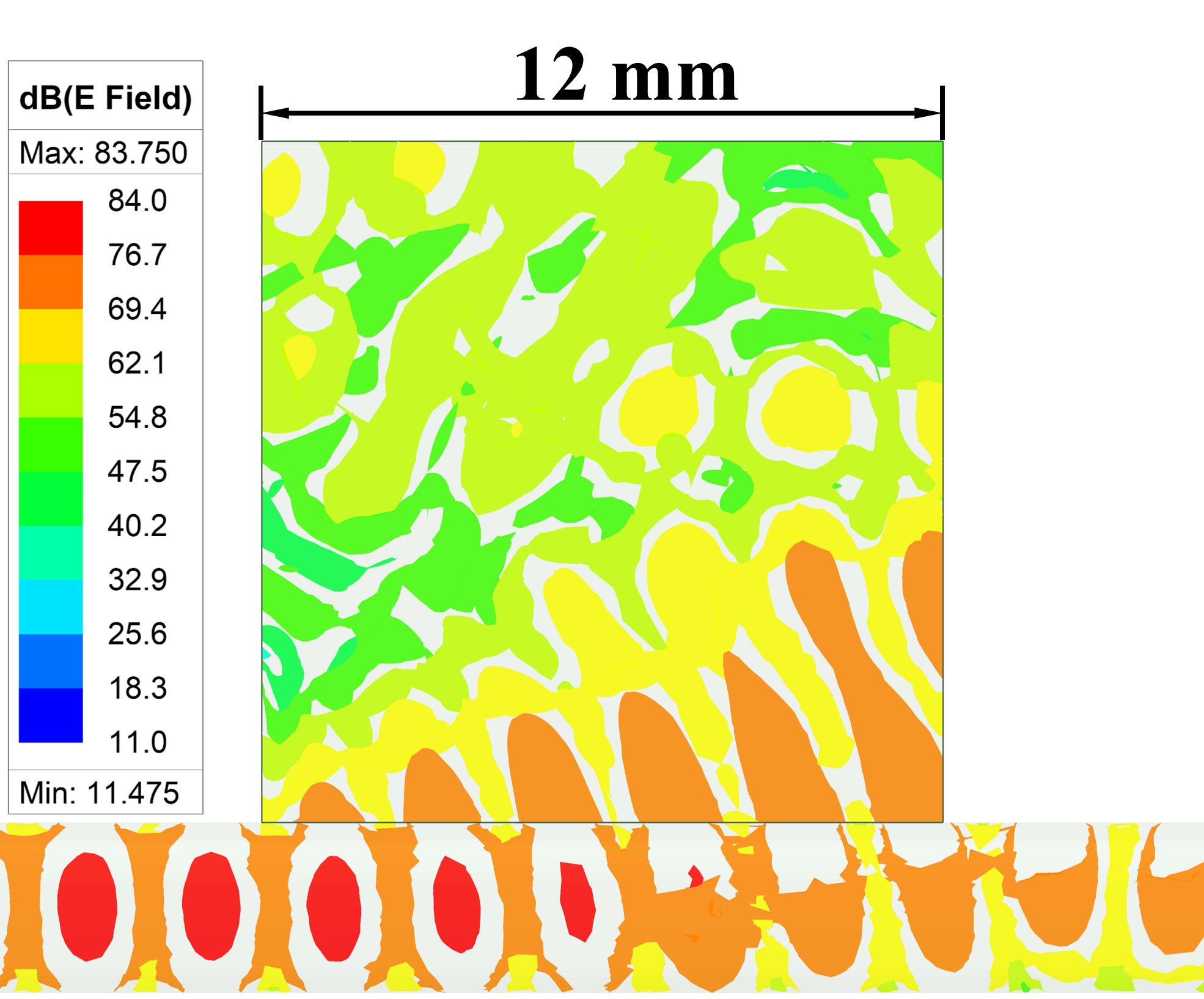}}\label{square.Fig}}
        \hfil
        \subfloat[]{\makebox[0.24\textwidth][c]{\includegraphics[height=1.30in]{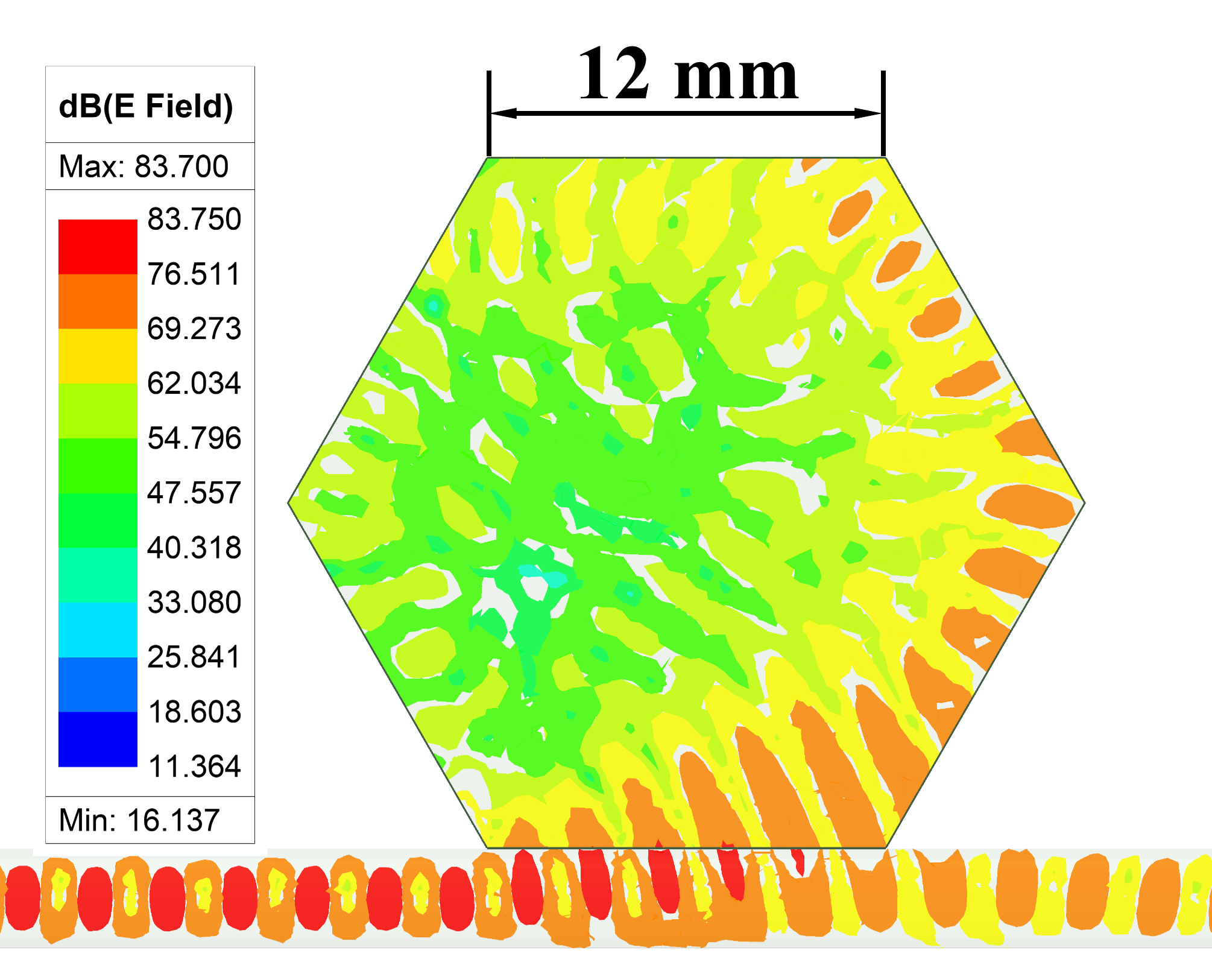}}\label{hexagon.Fig}}
        \hfil
        \subfloat[]{\makebox[0.24\textwidth][c]{\includegraphics[height=1.30in]{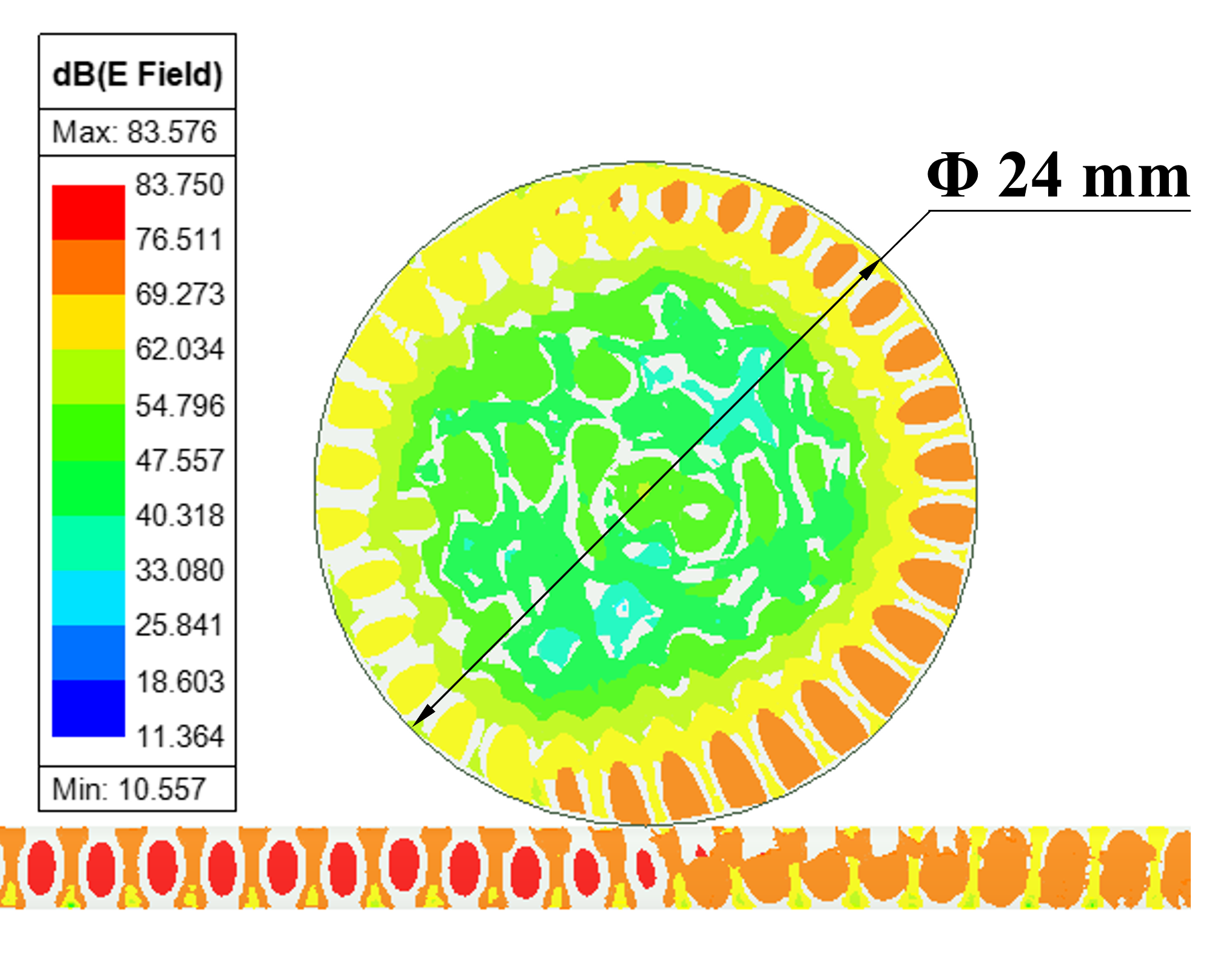}}\label{circle.Fig}}
        \\
        % Row 2: 2D antenna gain heat maps
        \subfloat[]{\includegraphics[width=0.24\textwidth]{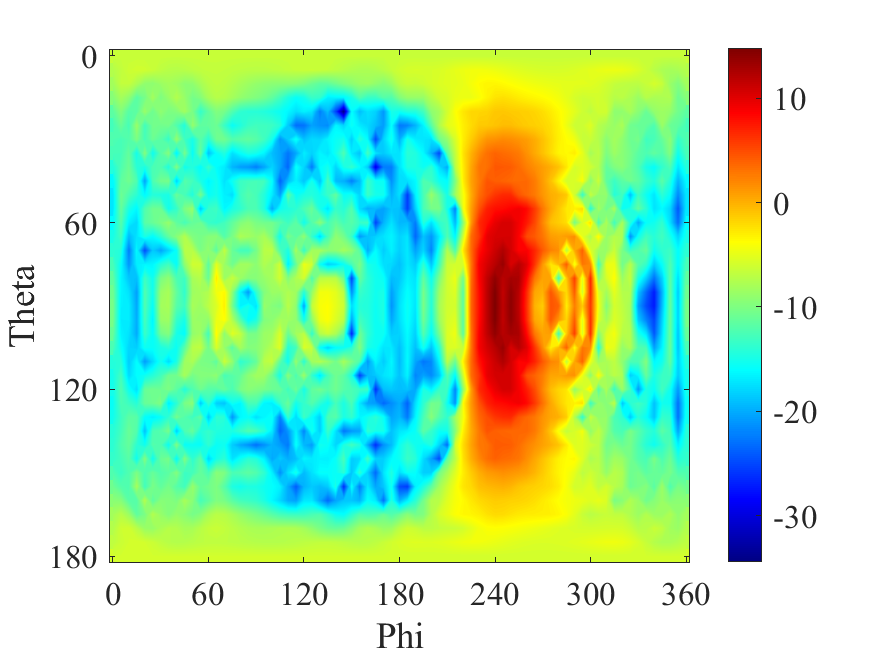}\label{tri.gain}}
        \hfil
        \subfloat[]{\includegraphics[width=0.24\textwidth]{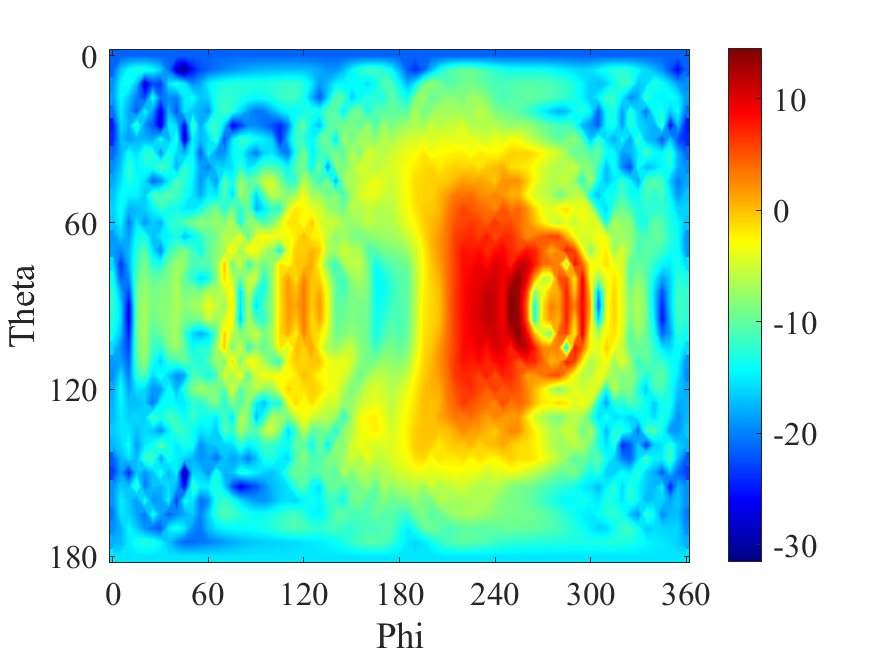}\label{square.gain}}
        \hfil
        \subfloat[]{\includegraphics[width=0.24\textwidth]{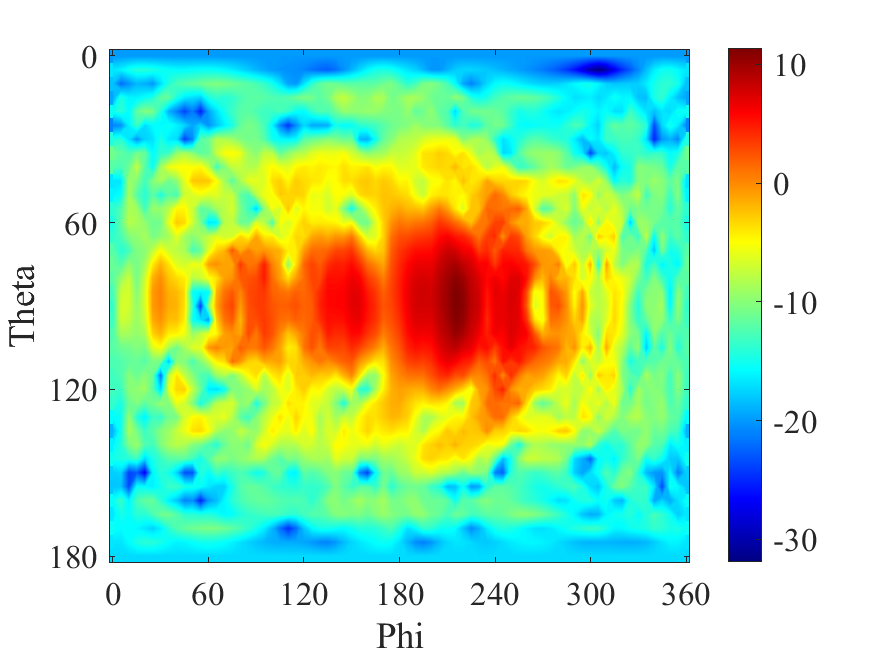}\label{hexagon.gain}}
        \hfil
        \subfloat[]{\includegraphics[width=0.24\textwidth]{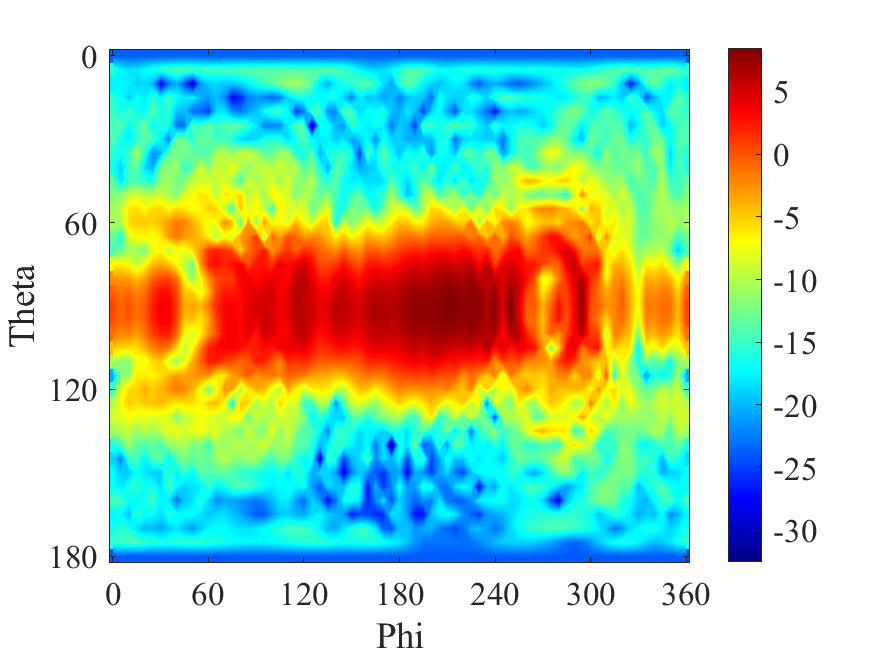}\label{circle.gain}}
        \\
        % Row 3: User-plane antenna gain distributions
        \subfloat[]{\includegraphics[width=0.24\textwidth]{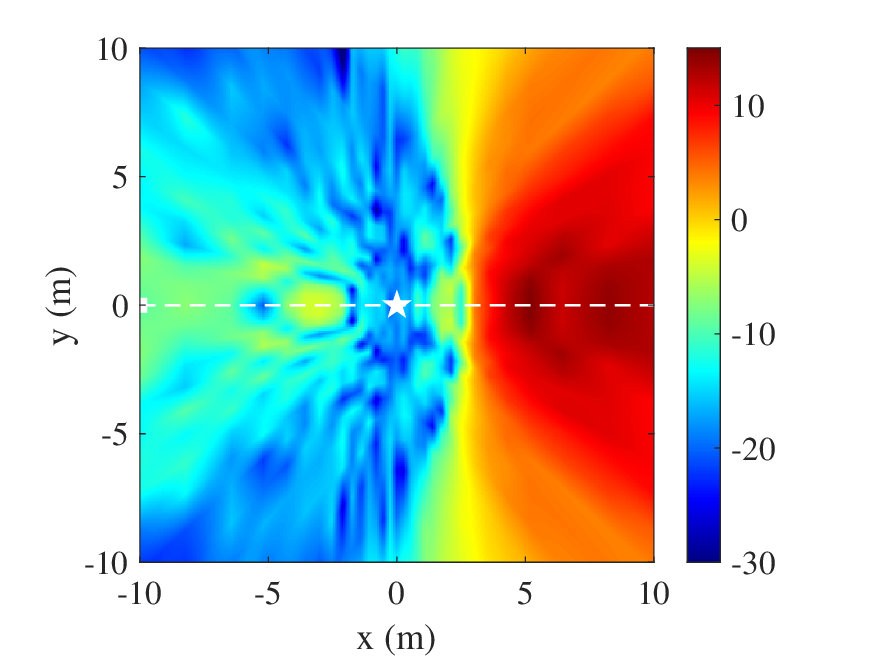}\label{tri.xy.gain}}
        \hfil
        \subfloat[]{\includegraphics[width=0.24\textwidth]{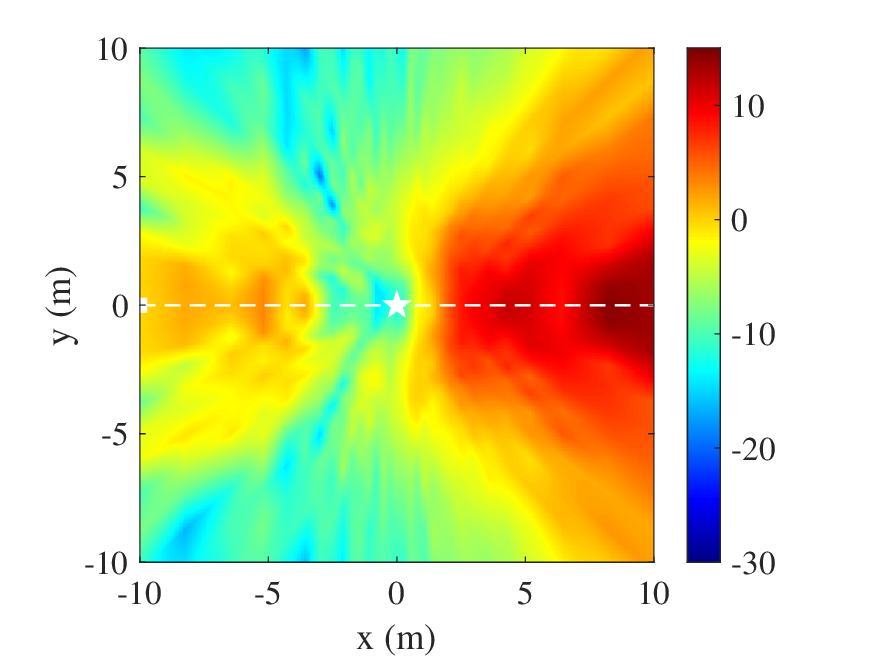}\label{square.xy.gain}}
        \hfil
        \subfloat[]{\includegraphics[width=0.24\textwidth]{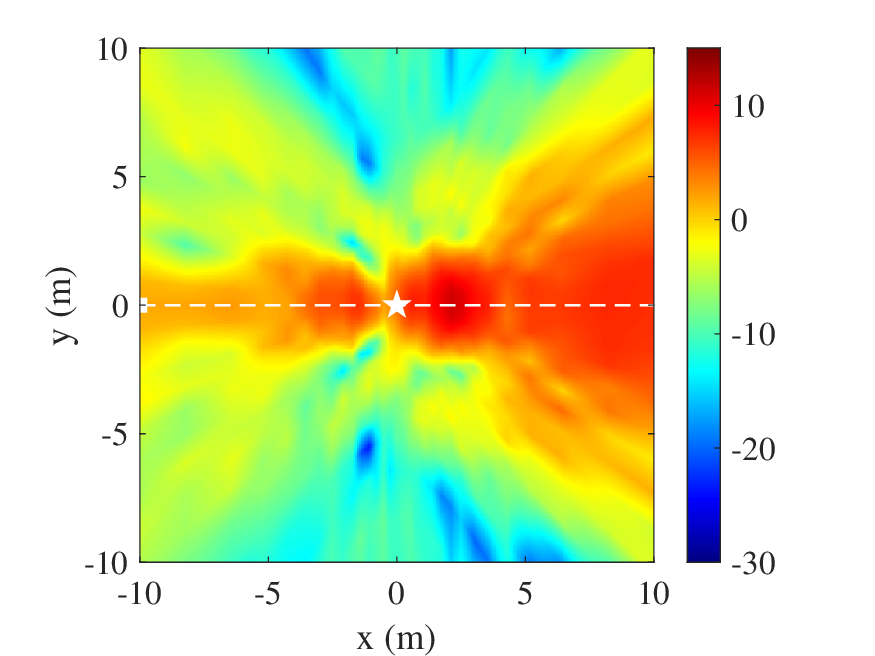}\label{hexagon.xy.gain}}
        \hfil
        \subfloat[]{\includegraphics[width=0.24\textwidth]{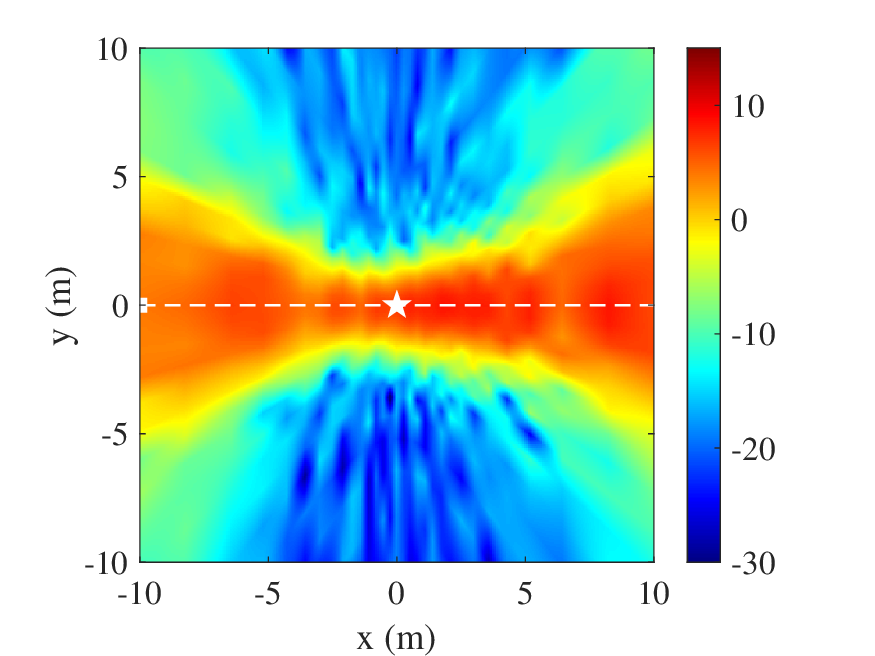}\label{circle.xy.gain}}
        \caption{Electric field distributions, 2D antenna gain heat maps, and user-plane antenna gain distributions for PAs with different geometries. (a), (e), (i) Triangle PA. (b), (f), (j) Square PA \cite{li2026geometry}. (c), (g), (k) Hexagon PA. (d), (h), (l) Circle PA.}\label{geometry.Fig}
\end{figure*}

% From left to right, the number of polygon edges increases from 3 (triangle) to $\infty$ (circle), illustrating the transition from strong directivity to near-isotropic radiation.

To highlight the role of PA geometry, several representative configurations are examined. Using the High Frequency Structure Simulator (HFSS), we simulate a flat PA structure and present its internal electric field distribution to demonstrate how geometric parameters modify the far-field radiation pattern at 60 GHz. Simulation results reveal that the electric field transitions gradually from the waveguide into the PA at an obtuse angle. 

As shown in Fig. \ref{tri.Fig}, the regular triangular PA's aperture orientation is nearly orthogonal to the field propagation direction, producing pronounced radiation and a distinct main lobe. As illustrated in Fig. \ref{square.Fig}, the geometry of the PAs significantly influences the boundary conditions, thereby altering the electromagnetic field distribution within the structure. Specifically, the square PA produces a more distributed internal field with several strong-field regions, whereas the triangular PA concentrates the field along a dominant propagation path, leading to a more pronounced main radiation lobe. The hexagonal and circular cases further show how smoother PA boundaries modify the internal field propagation. As depicted in Figs. \ref{hexagon.Fig} and \ref{circle.Fig}, variations in the edge angles of the PA effectively alter the propagation direction of the electric field. Specifically, the arc-like edges guide the field to propagate along the curved boundary of the PA, redirecting the energy flow around the antenna structure. To provide a concrete example of $G_{\mathcal{G}}(\theta,\phi)$, consider the in-plane cut at $\theta = 90^\circ $, corresponding to the plane containing the waveguide and the PA. The regular triangular PA exhibits a peak gain of 14.80 dB at $\phi = 240^\circ$, indicating a pronounced directional main lobe. By contrast, the circular PA provides a broader angular response, with its gain ranging from 4.16 to 8.46 dB over $\phi \in [114^\circ, 250^\circ]$. To further illustrate the impact of PA geometry on user-plane coverage, Figs. \ref{tri.xy.gain}--\ref{circle.xy.gain} present the 2D gain distributions in the user plane at $z = 0$, where $L = 20$ m and $d = 3$ m. The triangular PA concentrates its gain toward one side of the user plane, whereas the circular PA provides a broader and more uniform gain distribution along the waveguide axis.

These geometry-dependent radiation characteristics have important implications for PASS design. Most existing PASS studies exploit PA placement to mitigate free-space path loss and activate multiple PAs to realize beamforming. In contrast, PA geometry introduces an additional design degree of freedom. For example, a highly directional PA, such as the triangular configuration, can concentrate radiated power toward a prescribed service area while suppressing unnecessary leakage. By comparison, the circular PA provides a broader and more uniform azimuthal response within the PA plane, making it more suitable for local coverage when the users' directions are uncertain. Therefore, PA geometry should be optimized according to the system-level design objective.

\section{PA Rotation and Directional Control}
\label{sec:direction_control}
The geometry-dependent results above confirm that PA structure can reshape the radiation pattern. This section turns to direction control, where a designed PA structure is rotated or installed with a specific orientation to translate geometry-dependent radiation into steerable coverage.

\subsection{Directional Steering by Arc PAs}

\begin{figure*} [t!]
        \centering
        % Row 1: E-field distributions for arc PAs (use bounding box to normalize visual size)
        \subfloat[]{\includegraphics[height=1.00in]{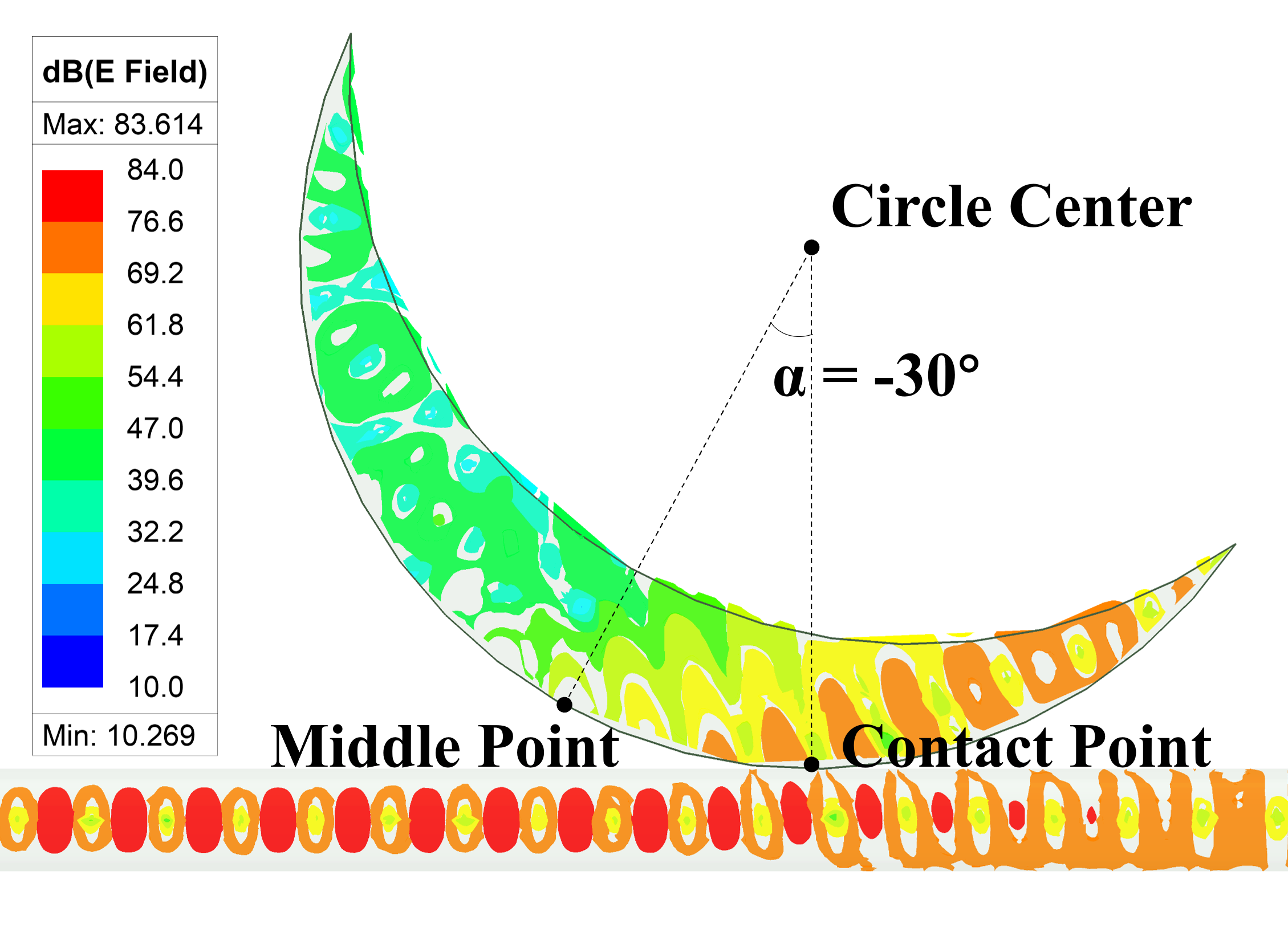}\label{arc_m30.Fig}}
        \hfil
        \subfloat[]{\includegraphics[height=0.97in]{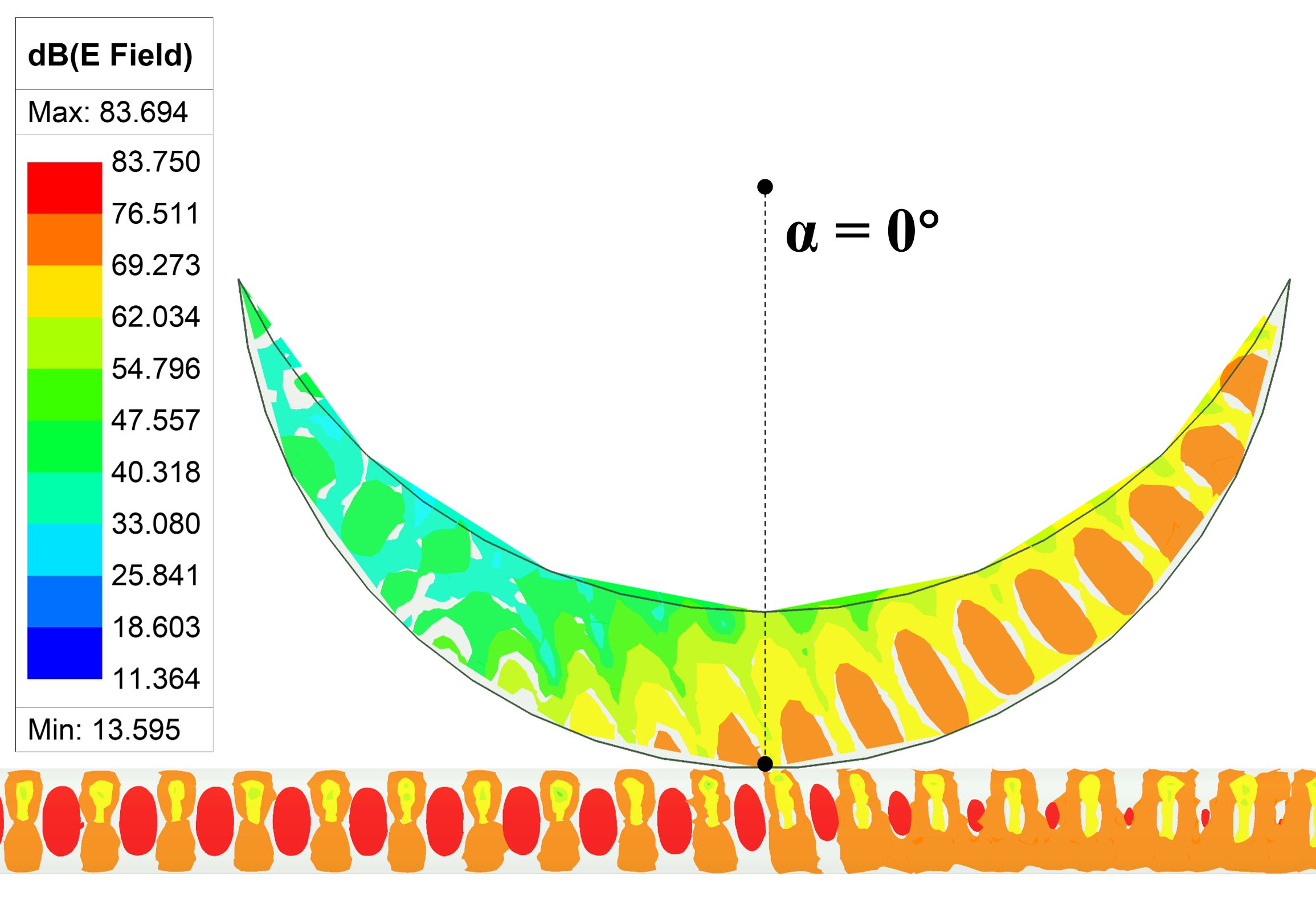}\label{arc_0.Fig}}
        \hfil
        \subfloat[]{\includegraphics[height=0.99in]{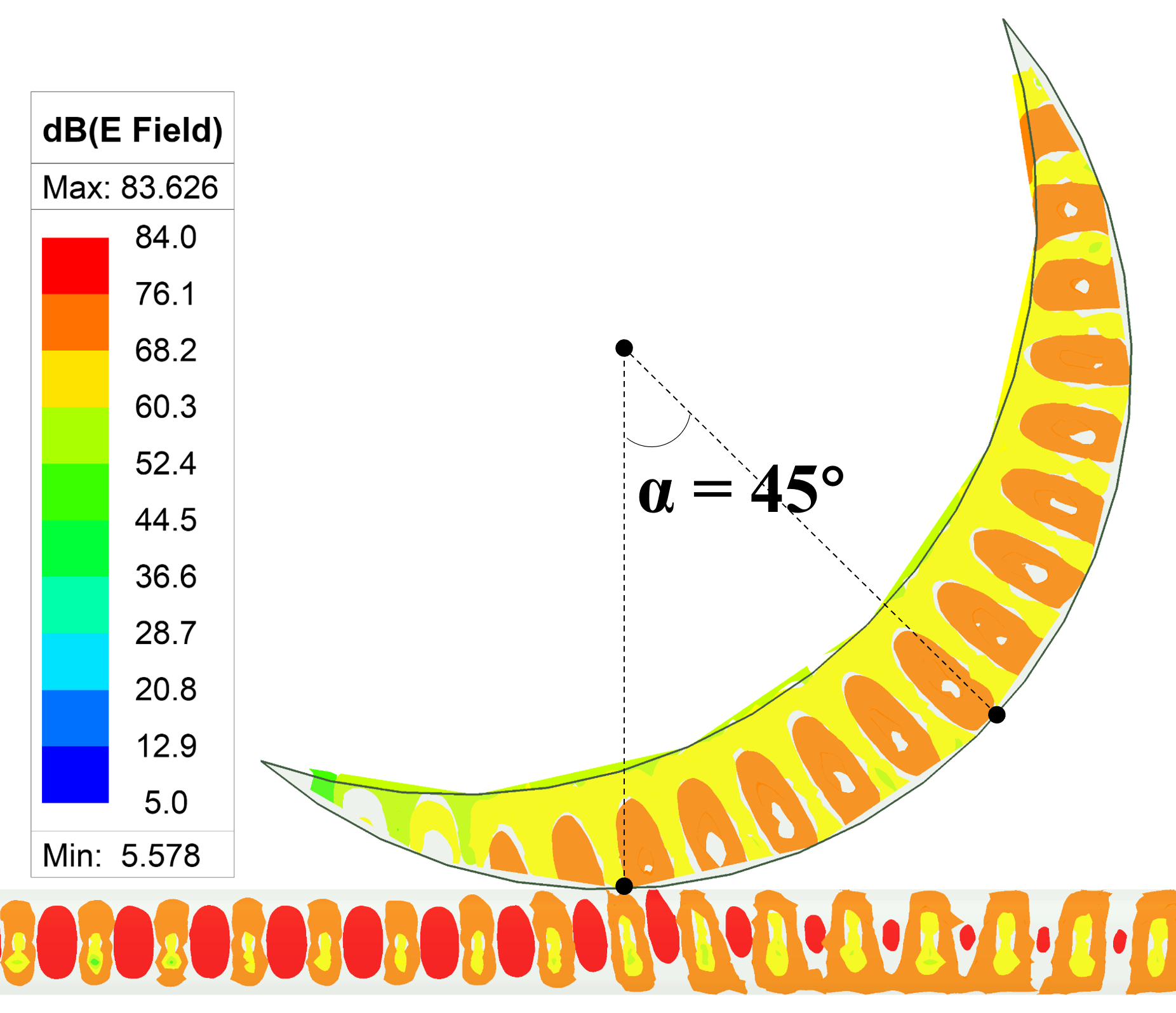}\label{arc_45.Fig}}
        \hfil
        \subfloat[]{\includegraphics[height=0.95in]{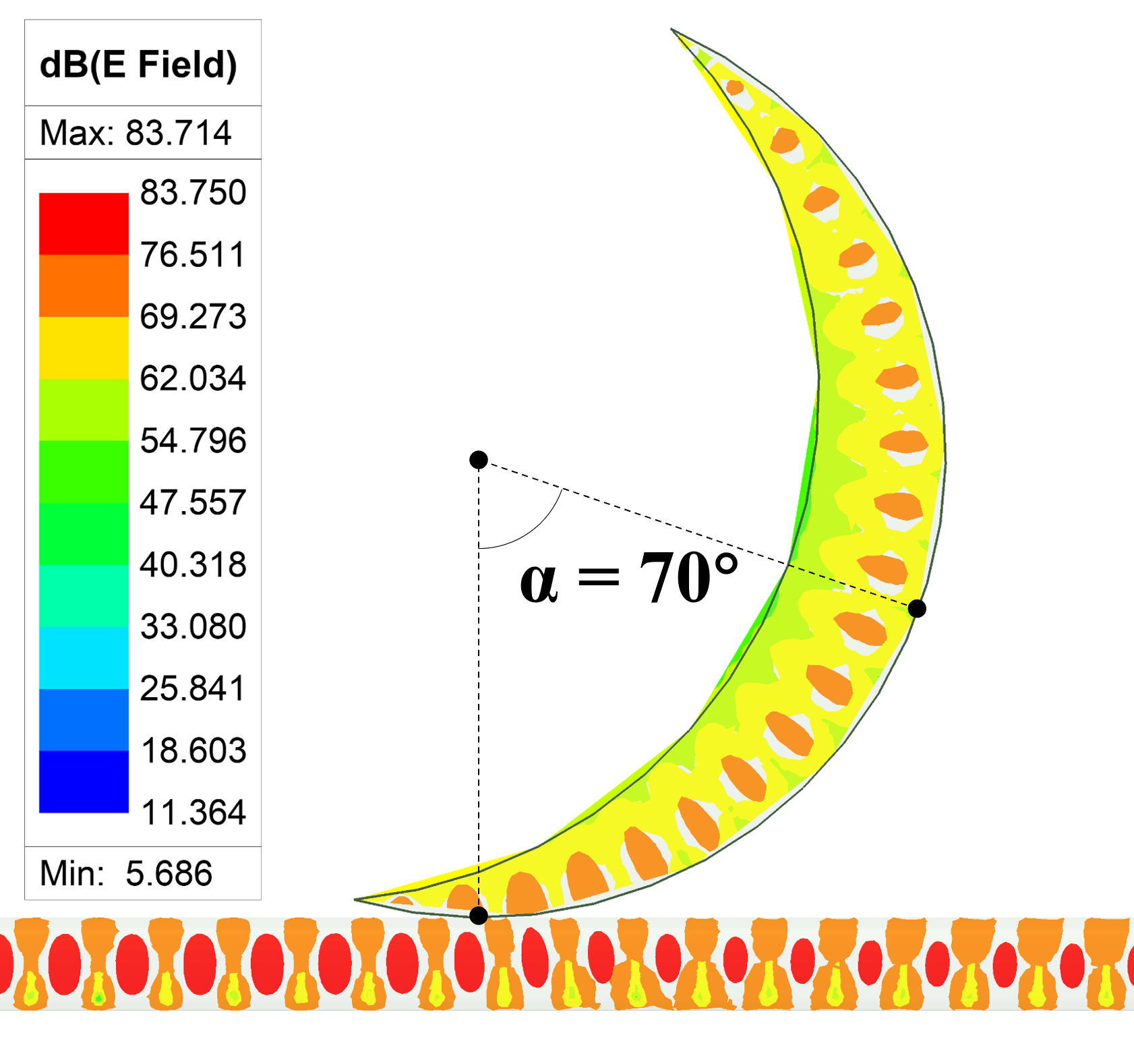}\label{arc_70.Fig}}
        \\
        % Row 2: 2D antenna gain heat maps for arc PAs
        \subfloat[]{\includegraphics[width=0.20\textwidth]{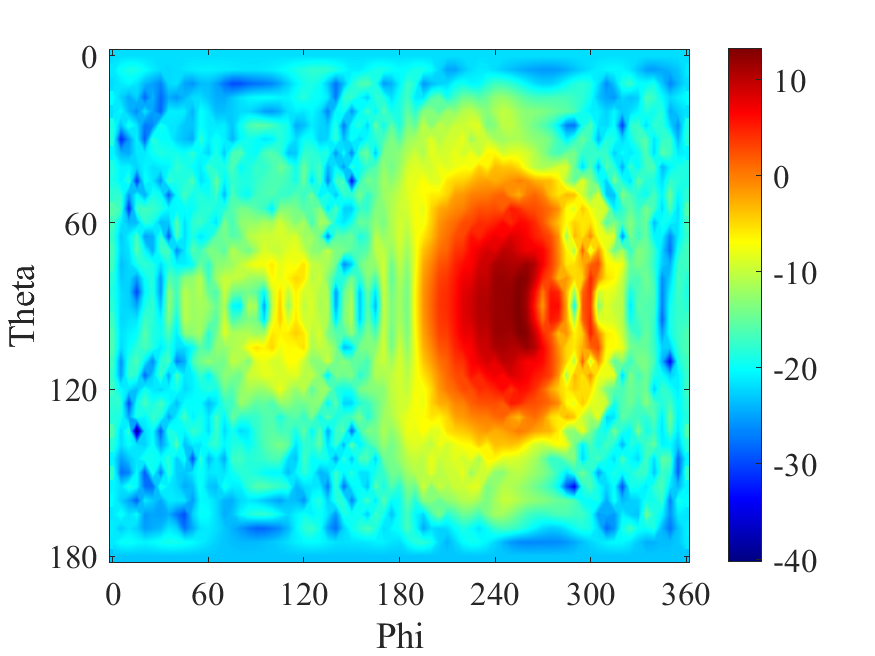}\label{arc_m30.gain}}
        \hfil
        \subfloat[]{\includegraphics[width=0.20\textwidth]{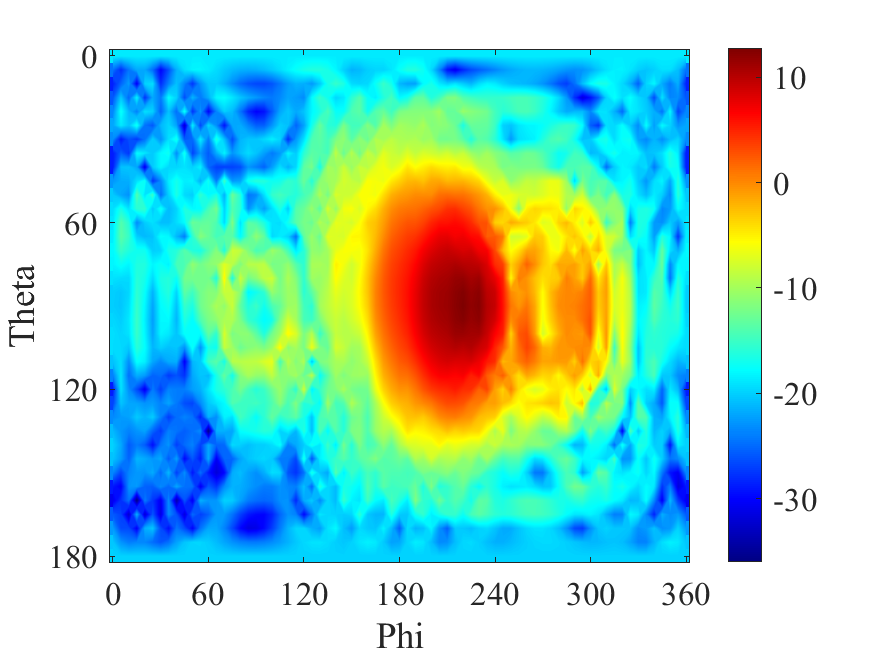}\label{arc_0.gain}}
        \hfil
        \subfloat[]{\includegraphics[width=0.20\textwidth]{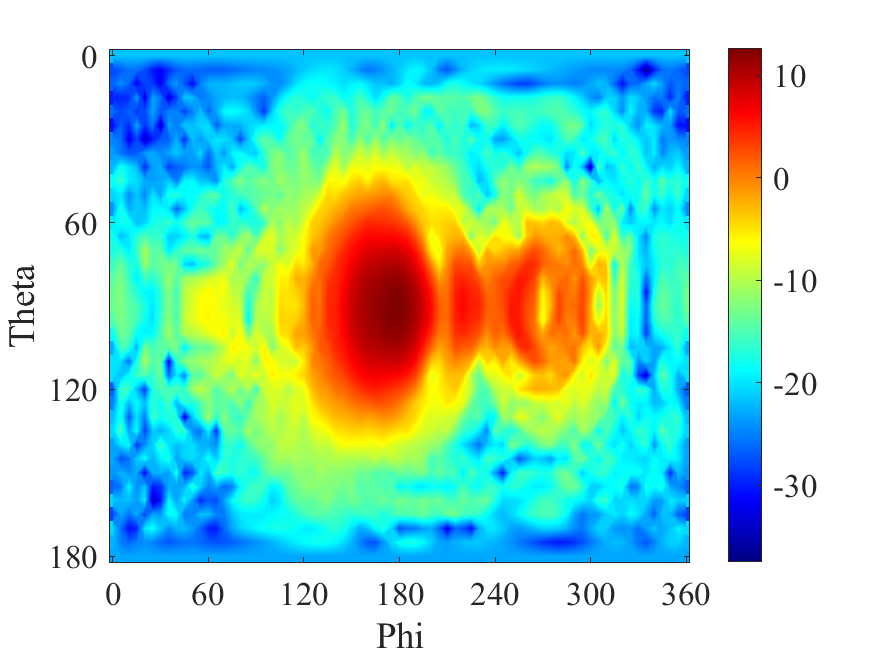}\label{arc_45.gain}}
        \hfil
        \subfloat[]{\includegraphics[width=0.20\textwidth]{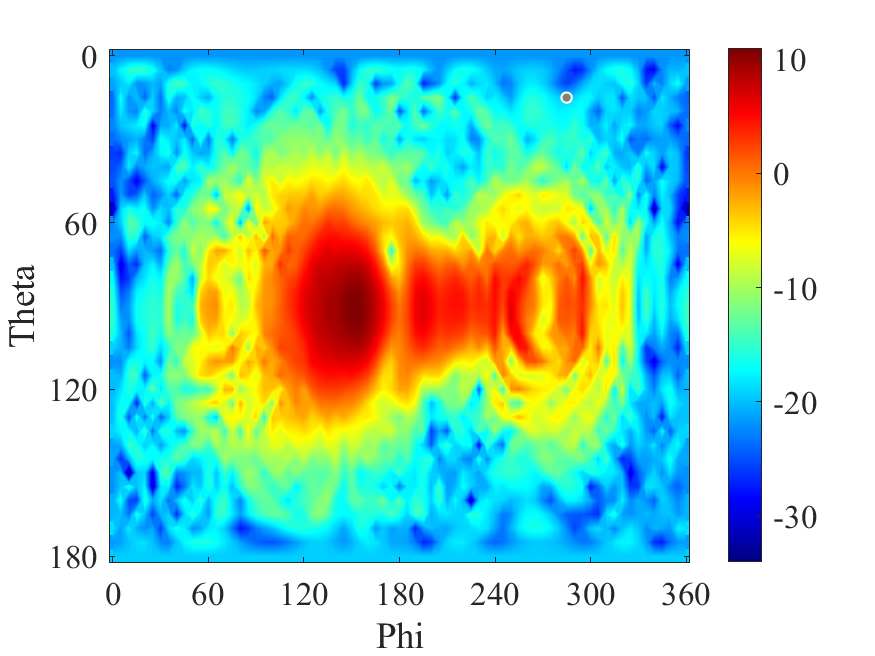}\label{arc_70.gain}}
        \\
        % Row 3: User-plane antenna gain distributions for arc PAs
        \subfloat[]{\includegraphics[width=0.20\textwidth]{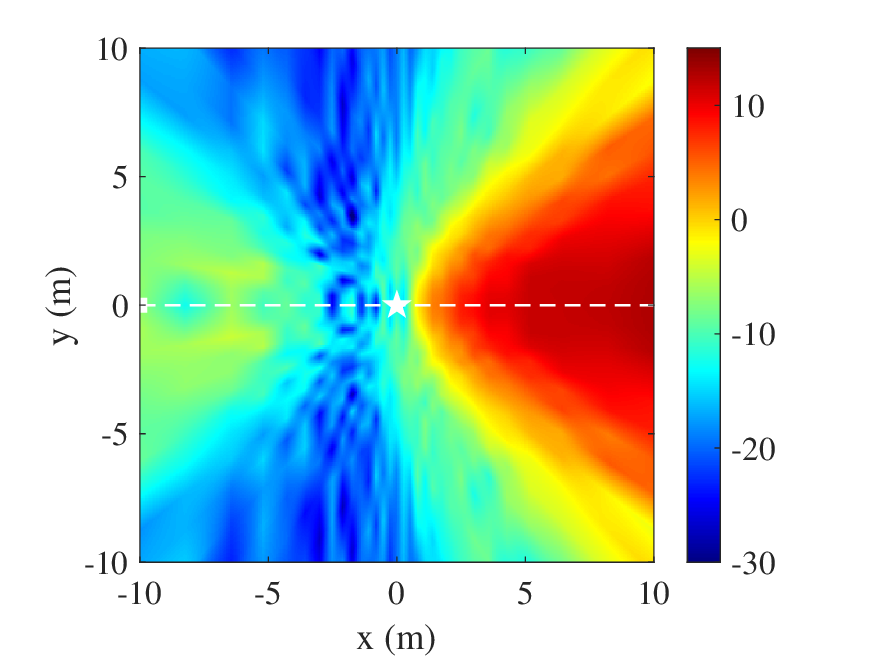}\label{arc_m30.xy.gain}}
        \hfil
        \subfloat[]{\includegraphics[width=0.20\textwidth]{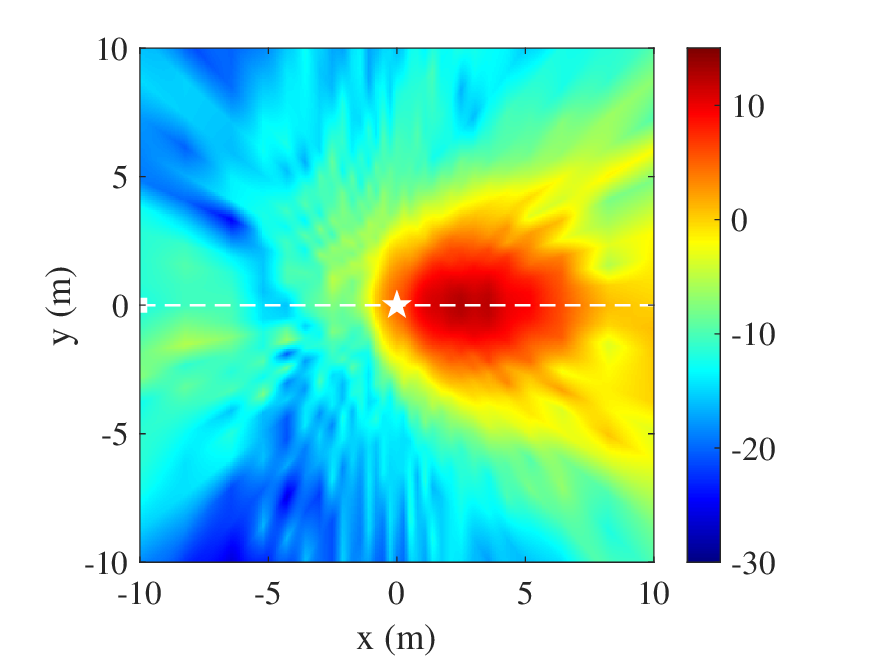}\label{arc_0.xy.gain}}
        \hfil
        \subfloat[]{\includegraphics[width=0.20\textwidth]{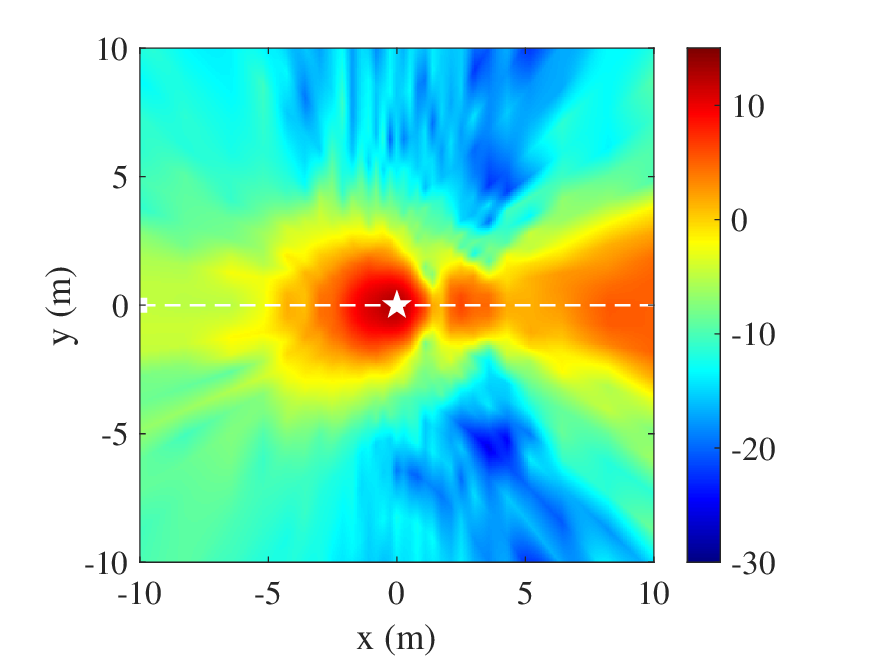}\label{arc_45.xy.gain}}
        \hfil
        \subfloat[]{\includegraphics[width=0.20\textwidth]{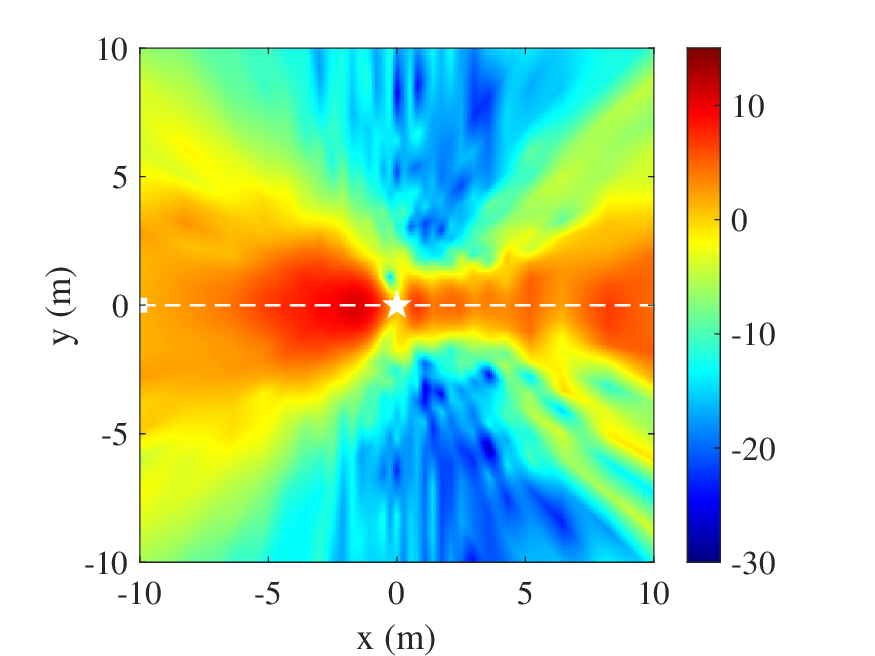}\label{arc_70.xy.gain}}
        \caption{Directional radiation steering by arc PAs. (a)--(l) Electric field distributions, 2D antenna gain heat maps, and user-plane antenna gain distributions for arc PAs with different arc rotation angles. (a), (e), (i) $\alpha = -30^{\circ}$; (b), (f), (j) $\alpha = 0^{\circ}$; (c), (g), (k) $\alpha = 45^{\circ}$; and (d), (h), (l) $\alpha = 70^{\circ}$.}\label{direction_control.Fig}
\end{figure*}

\begin{figure*} [t!]
        \centering
        % Orientation-based direction control
        \subfloat[]{\includegraphics[height=0.78in]{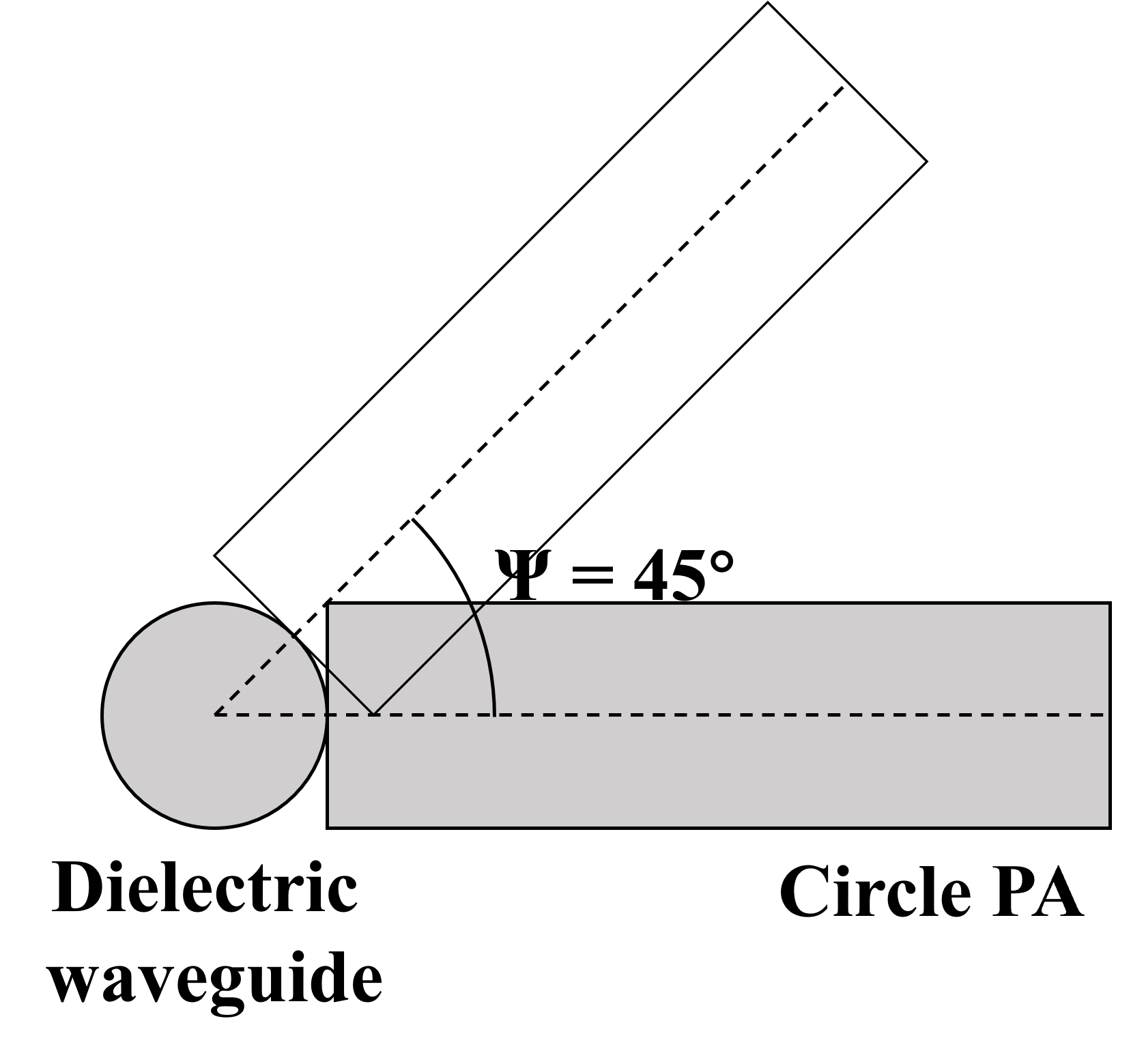}\label{rotate_schematic.Fig}}
        \hfil
        \subfloat[]{\includegraphics[height=0.78in]{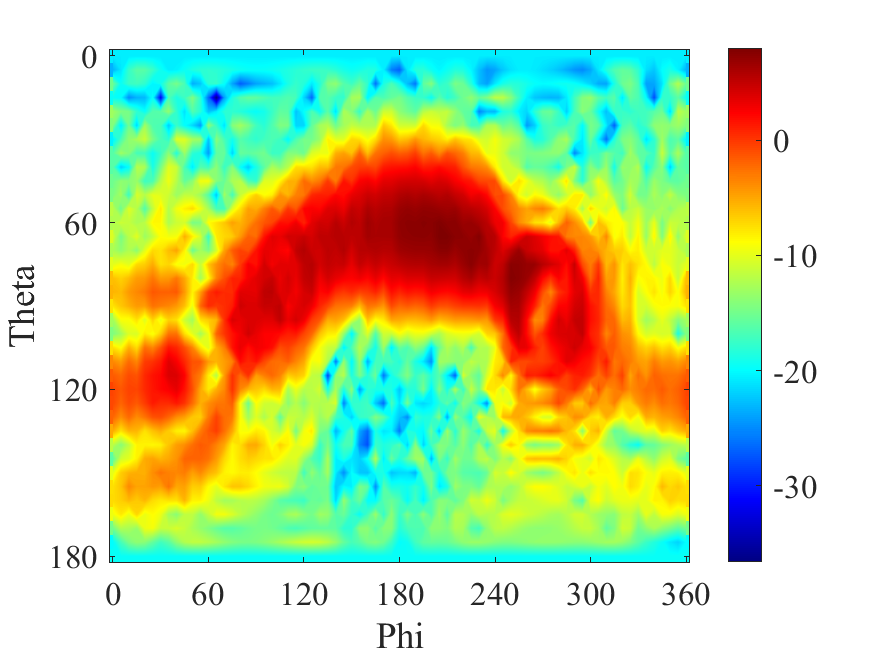}\label{rotate30.Fig}}
        \hfil
        \subfloat[]{\includegraphics[height=0.78in]{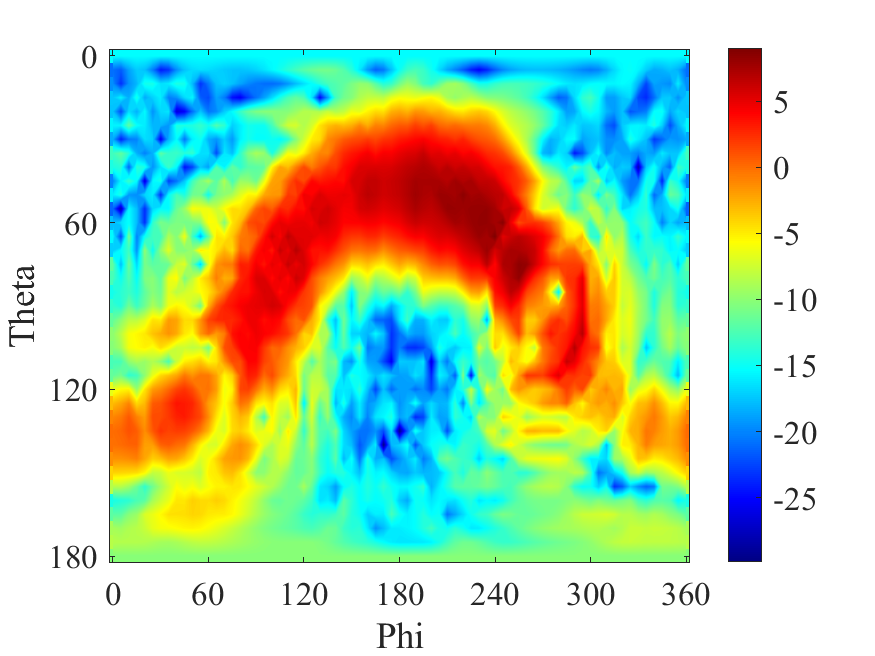}\label{rotate45.Fig}}
        \hfil
        \subfloat[]{\includegraphics[height=0.78in]{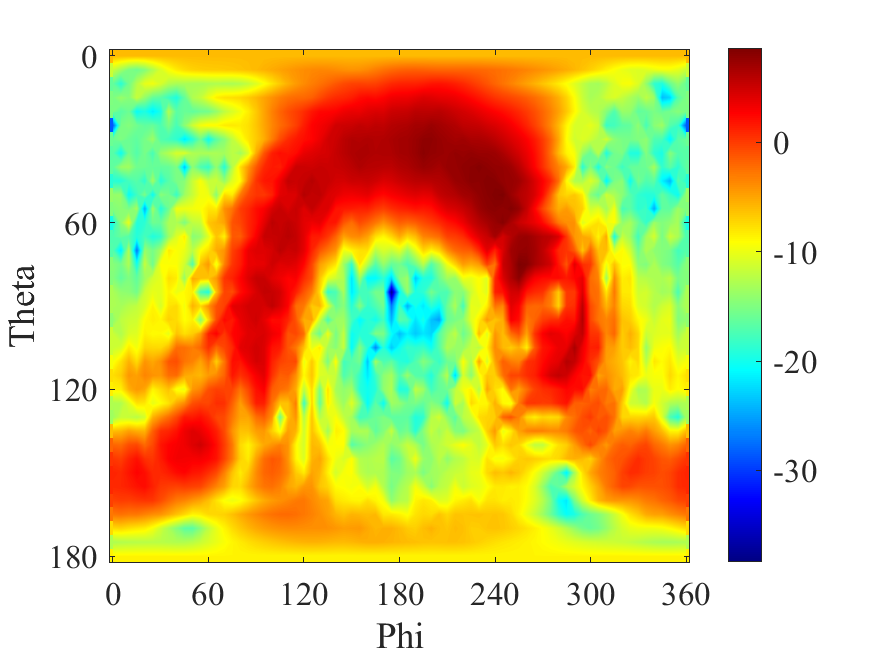}\label{rotate60.Fig}}
        \hfil
        \subfloat[]{\includegraphics[height=0.78in]{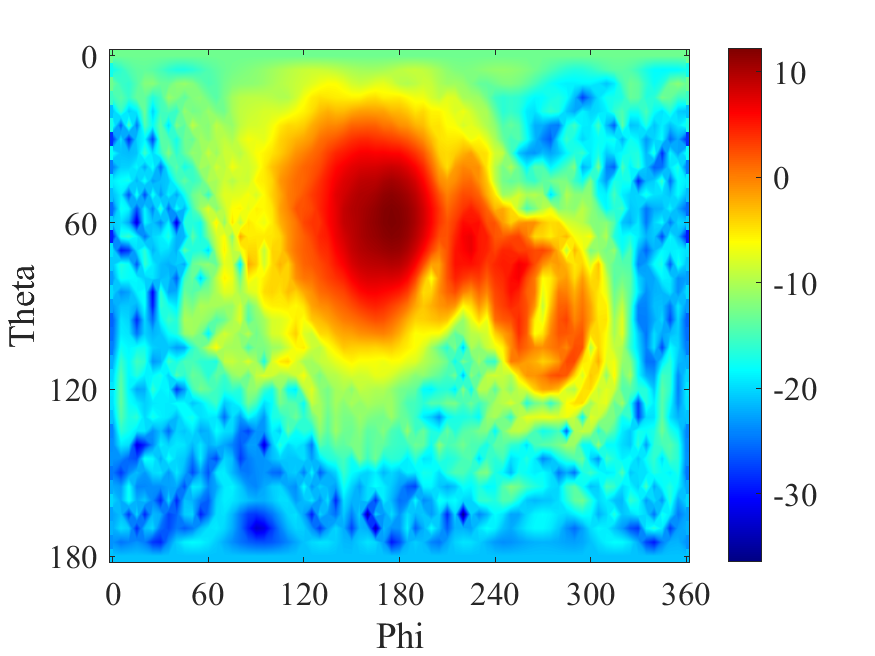}\label{arc_rotate.Fig}}
        \\
        \vspace{0.03in}
        % Row 2: User-plane antenna gain distributions under orientation control
        \subfloat[]{\includegraphics[width=0.20\textwidth]{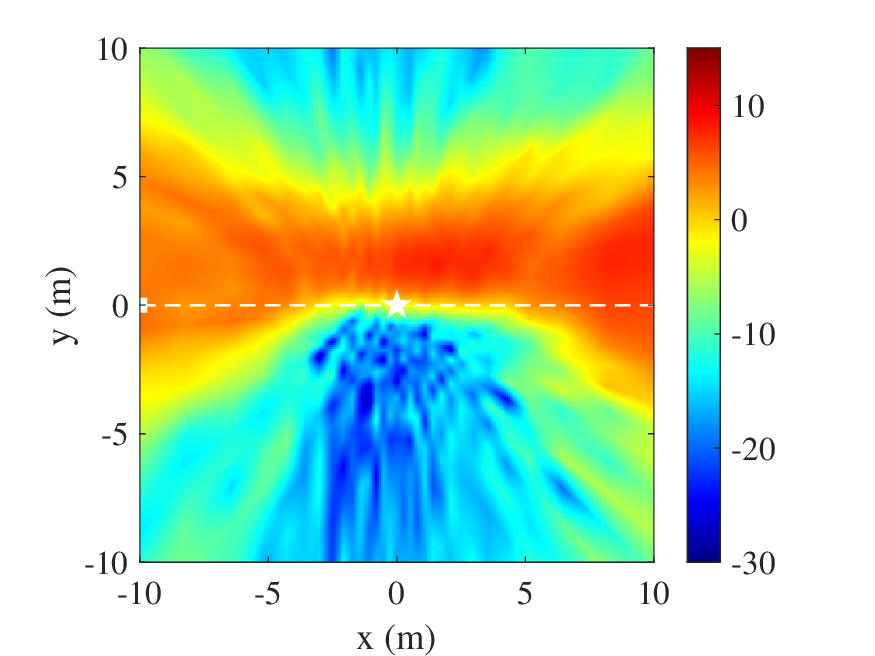}\label{rotate30.xy.gain}}
        \hfil
        \subfloat[]{\includegraphics[width=0.20\textwidth]{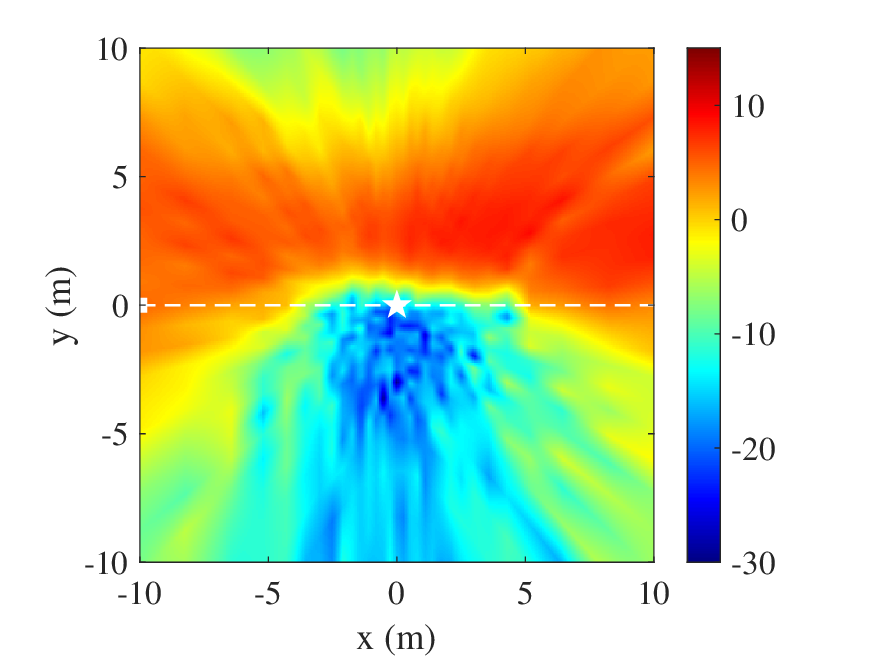}\label{rotate45.xy.gain}}
        \hfil
        \subfloat[]{\includegraphics[width=0.20\textwidth]{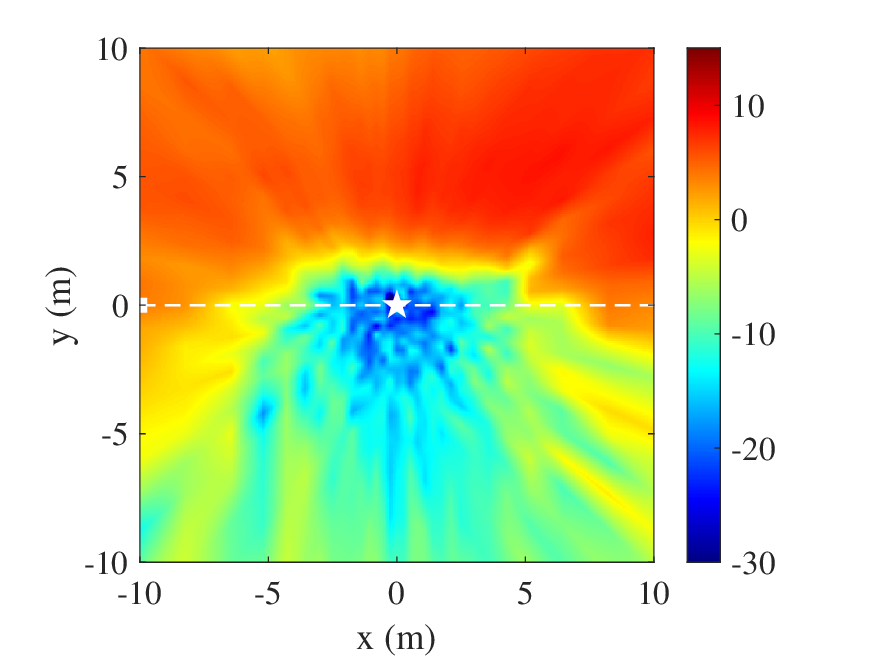}\label{rotate60.xy.gain}}
        \hfil
        \subfloat[]{\includegraphics[width=0.20\textwidth]{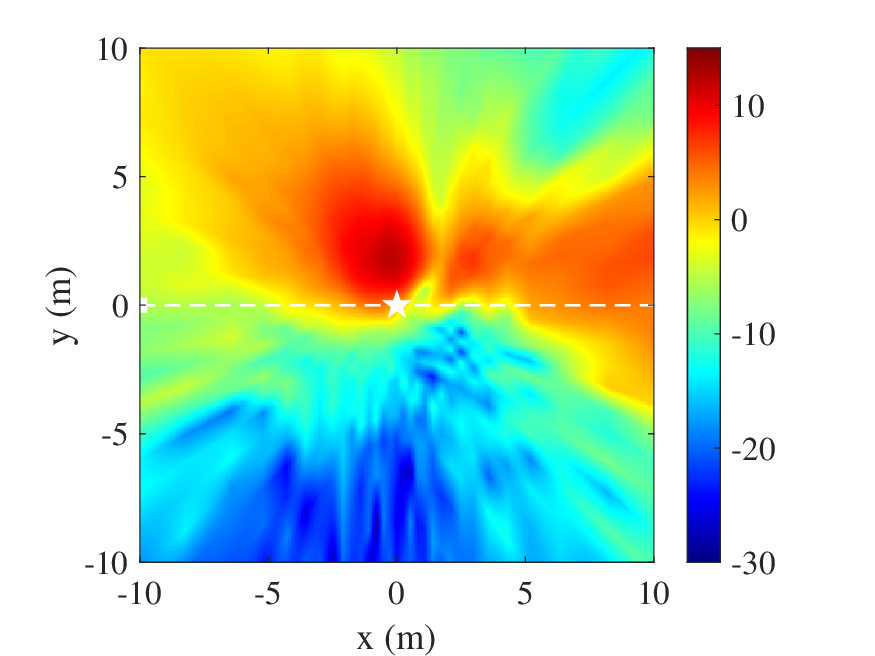}\label{arc_rotate.xy.gain}}
        \caption{Orientation-based radiation direction control. (a) Schematic of the rotated PA on the dielectric waveguide, where $\Psi$ denotes the orientation angle. (b)--(d) Far-field gain patterns of a circular PA for $\Psi = 30^{\circ}$, $45^{\circ}$, and $60^{\circ}$, respectively. (e) Far-field gain pattern of an arc PA with $\alpha = 45^{\circ}$ and $\Psi = 45^{\circ}$, illustrating the joint effect of PA shape and orientation. (f)--(i) User-plane gain distributions corresponding to (b)--(e), respectively.}\label{orientation_control.Fig}
\end{figure*}

An arc-shaped PA provides a direct way to make the radiation direction tunable through geometry \cite{li2026geometry}. When the arc is rotated relative to the waveguide, the strongly excited part of the arc moves to different angular sectors, and the main lobe can follow this change. This is because the far-field pattern is governed by the electric field distribution along the radiating edges. In conventional PA geometries, the coupled field may weaken across the dielectric block, making the opposite edges less effective in forming strong lobes. The arc PA instead keeps the radiating edge close to the strongly excited region, allowing a stronger field to be maintained along the arc. 

To demonstrate the steering capability, we define a rotation angle $\alpha$ that parameterizes the orientation of the arc relative to the waveguide. Figs.~\ref{arc_m30.Fig}--\ref{arc_70.xy.gain} show the internal electric field distributions, corresponding 2D antenna gain heat maps, and user-plane gain distributions for four representative configurations: $\alpha = -30^{\circ}$, $0^{\circ}$, $45^{\circ}$, and $70^{\circ}$. As $\alpha$ changes, the main lobe rotates accordingly. Specifically, the peak gain angles are approximately $254^{\circ}$, $220^{\circ}$, $180^{\circ}$, and $150^{\circ}$ for $\alpha = -30^{\circ}$, $0^{\circ}$, $45^{\circ}$, and $70^{\circ}$, respectively, with half-power angular sectors $234^{\circ}$--$260^{\circ}$, $196^{\circ}$--$238^{\circ}$, $154^{\circ}$--$192^{\circ}$, and $130^{\circ}$--$166^{\circ}$. The user-plane gain maps show that varying the arc rotation angle progressively shifts the high-gain region, enabling controllable coverage steering across the service area. These results indicate that the arc PA provides a continuous degree of freedom for radiation steering through geometric reconfiguration alone.

\subsection{Orientation-Based Direction Control}

The preceding simulations reveal that PA radiation is predominantly concentrated within the plane of the PA structure. In practical deployments, however, users are distributed across three-dimensional space and are not necessarily located in the same plane as the PA, which may degrade the received signal strength. To extend coverage, an orientation angle $\Psi$ is introduced, as illustrated in Fig.~\ref{rotate_schematic.Fig}, which characterizes the rotation of the PA relative to the waveguide axis. Taking the circular PA as an example, the gain pattern in Fig.~\ref{circle.gain} provides the baseline case with $\Psi = 0^{\circ}$. Starting from this baseline, increasing $\Psi$ tilts the radiation pattern away from the original PA plane, as shown in Figs.~\ref{rotate30.Fig}--\ref{rotate60.Fig}. The corresponding user-plane gain distributions in Figs.~\ref{rotate30.xy.gain}--\ref{arc_rotate.xy.gain} further show that varying the PA orientation redirects the high-gain region, while joint shape-and-orientation control enables more flexible coverage steering.

For practical deployment, this is particularly attractive because PA orientation can be adjusted during installation without modifying the antenna geometry itself, providing a low-cost means of customizing coverage. Conventional MIMO beam steering usually relies on coordinated phase and amplitude control across multiple RF-fed antenna elements. Existing PASS beamforming methods also exploit hardware reconfigurability. In a single-waveguide PASS, activating a PA near the user mainly reduces the free-space propagation distance, while activating multiple PAs enables pinching beamforming by aligning the phases of signals radiated from different waveguide positions. In multi-waveguide PASS, different waveguides can be fed by different RF chains, enabling transmit precoding or digital beamforming together with PA placement. These approaches are effective, but they often require multiple radiation points, accurate phase alignment, or additional waveguides. Geometry and orientation control provide a complementary route by shaping the radiation pattern of each PA element itself. Therefore, for quasi-static coverage scenarios, PA geometry and orientation can provide initial direction control at the element level before multi-PA combining or digital beamforming is applied.

In this framework, shape determines the beam type, location governs path loss, and orientation steers the beam toward the intended service area. Furthermore, Fig.~\ref{arc_rotate.Fig} shows an arc PA with $\alpha = 45^{\circ}$ and $\Psi = 45^{\circ}$, demonstrating that shape and orientation can be jointly exploited for more flexible radiation direction control. By providing directionality at the element level, PA geometry and orientation may reduce the need for densely activated PAs, saving waveguide space and simplifying compact PASS implementations.

\begin{figure*} [t!]
        \centering
        \subfloat[]{\includegraphics[height=2.00in]{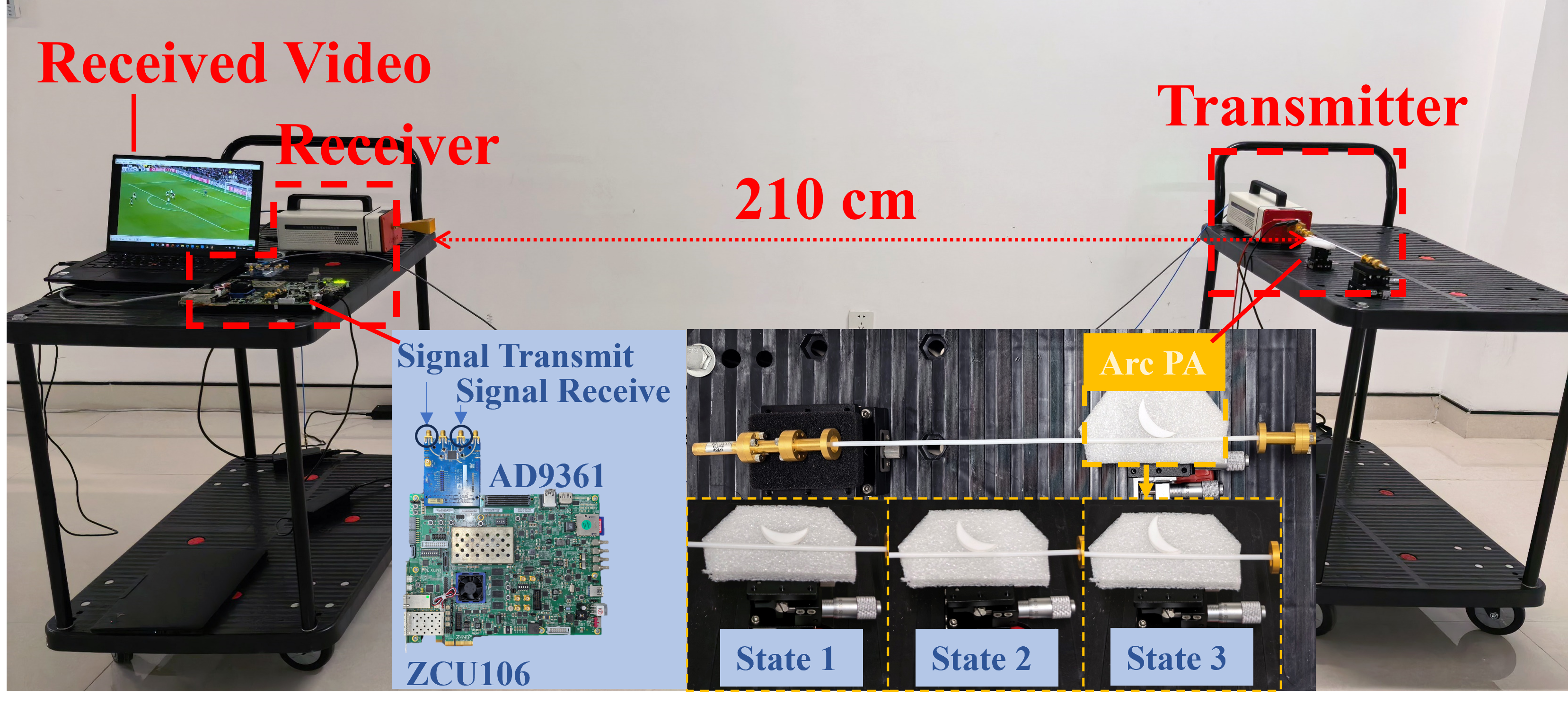}\label{demo_video.Fig}}
        \hfil
        \subfloat[]{\includegraphics[height=2.00in]{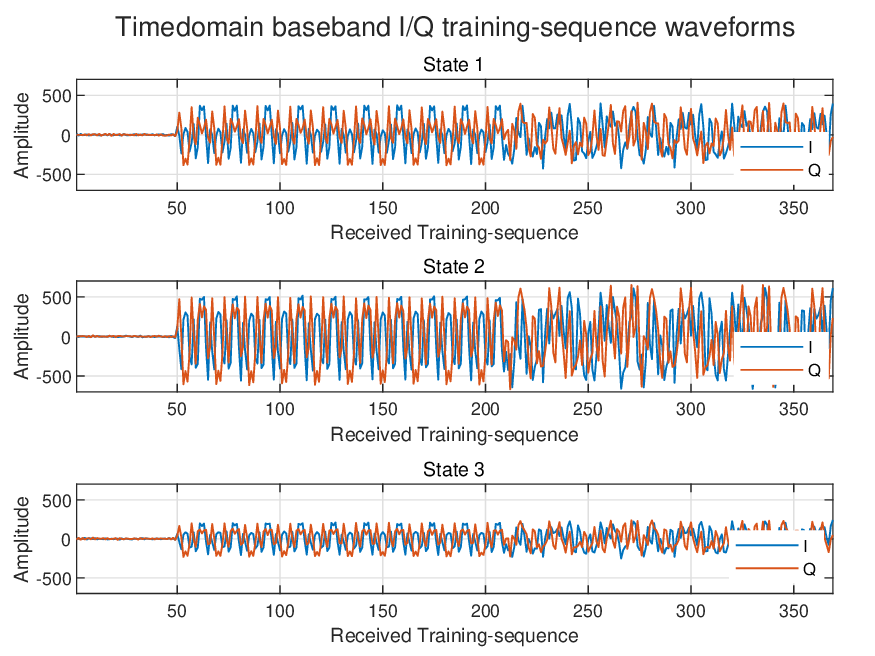}\label{demo_waveform.Fig}}
        \caption{Prototype-based communication demonstration using an arc PA. (a) Video setup with a transmitter, a receiver, and an arc PA attached to the dielectric waveguide. (b) Received baseband I/Q waveforms under different arc PA states.}\label{demo.Fig}
\end{figure*}

\subsection{Prototype Validation and Communication Demonstration}
The simulation results above show that PA shape and orientation can translate geometric changes into direction-dependent gain. To validate this effect, a prototype is built and tested through a video transmission experiment.

The video transmission experiment is shown in Fig.~\ref{demo_video.Fig}. In this system, the video data stream is generated by a field programmable gate array (FPGA) board using orthogonal frequency division multiplexing (OFDM), and processed by AD9361 and an up converter to 60 GHz radio frequency (RF) signals. The RF signals are then fed into the dielectric waveguide and radiated by the arc PA. A down converter with horn antenna receive the signal and send it to the receive port in the AD9361, then, FPGA decode the signal and use the ethernet send the data stream to a laptop which used as the monitor for the video data.

% \footnote{In the prototype demonstration, the laptop acts as both the video source and display terminal. The source video stream is delivered to the FPGA board via Ethernet, transmitted through the 60-GHz PASS link after baseband and RF processing, and displayed after receiver-side recovery. Thus, Fig.~\ref{demo_video.Fig} highlights the received-video display and wireless link setup.}

As illustrated in Fig.~\ref{demo_video.Fig}, the arc PA is adjusted among three representative states while the transmitter, waveguide, and receiver remain fixed. This setup isolates the impact of PA rotation from other system factors and shows how the radiation direction affects the received signals. Fig.~\ref{demo_waveform.Fig} presents the received OFDM training pilots under different arc PA states. The visible change in pilot amplitude indicates that varying the rotation angle $\alpha$ changes how much radiated energy is directed toward the receiver, thereby affecting the received signal quality.

From a practical perspective, future PASS prototypes also require controllable mounting or actuation structures for PA activation, rotation, and orientation adjustment. These auxiliary structures cannot be treated as purely mechanical components, because dielectric or supporting materials placed near the waveguide may disturb the evanescent field and change the coupling behavior. Therefore, PA geometry should be co-designed with the control structure to preserve the desired radiation characteristics while enabling reliable reconfiguration. As PASS is still at an early stage of development, such geometry-control co-design remains an important open challenge for practical directional PA systems.

\begin{figure*} [t!]
        \centering
        \includegraphics[width=1\textwidth]{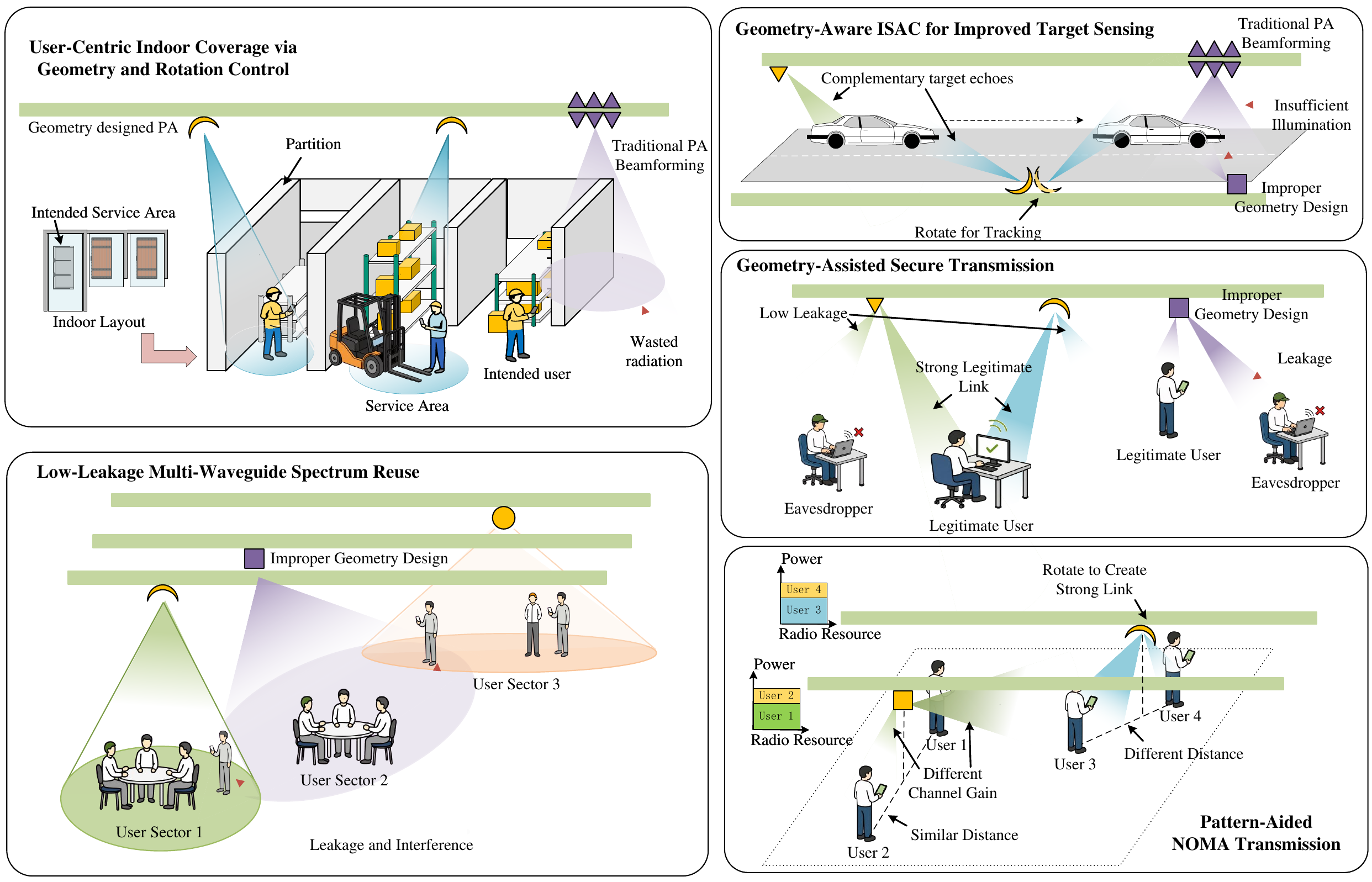}
        \caption{Promising applications of geometry-aware directional PA systems.}\label{app_scenarios.Fig}
\end{figure*}

\section{Promising Applications of Geometry-Aware Directional PA Systems}

The geometry-dependent radiation and direction-control capabilities discussed above create new opportunities for PASS beyond location tuning. Several promising application directions are illustrated in Fig.~\ref{app_scenarios.Fig} and discussed in the following subsections.

\subsection{User-Centric Indoor Coverage via Geometry and Rotation Control}

Indoor coverage is an important application scenario for PASS, where walls, shelves, corridors, and partitions often cause nonuniform propagation conditions. In a single-waveguide PASS, directionality can be roughly adjusted by coherently combining signals from several activated PAs through position and phase optimization \cite{ding2025flexible}. PA arrays can provide finer direction control, but they require more PA elements and waveguides, making them less suitable for compact indoor deployments. For indoor deployments, the waveguide may be short or interrupted by room boundaries, so only a few PAs can be installed or activated. Geometry and rotation control provide a compact approach to realizing user-centric coverage under these constraints. The PA shape determines the general radiation patterns, while rotation and installation orientation align the high-gain region with the intended users. By adapting PA geometry and rotation to user locations, PASS directs energy toward intended users while suppressing leakage. In this way, geometry-aware directional PAs actively reshape the local wireless propagation environment around the users, rather than merely adapting PA locations to a fixed propagation environment. This user-centric reconfiguration can reduce wasted radiation and improve coverage reliability in complex indoor spaces.

\subsection{Low-Leakage Multi-Waveguide Spectrum Reuse}

Multi-waveguide PASS can reuse the same time-frequency resources by allowing different waveguides to carry different user signals \cite{zhao2025pinching,li2026cognitive}. The main limitation is not only the distance between a PA and its intended user, but also the radiation leakage from one waveguide toward users served by other waveguides. Without geometry design, each activated PA may radiate energy into unnecessary directions, and these leaked components can become co-channel interference in free space. PA geometry design addresses this problem at the radiation source, since it determines how each PA distributes energy around the waveguide. By assigning suitable PA shapes and orientations to different waveguides, each signal can be directed toward its intended user with less leakage to others. In this way, multiple waveguides can support different signals in the same time-frequency resource with weaker interference.

\subsection{Geometry-Aware ISAC for Improved Target Sensing}

% PASS-assisted ISAC benefits from distributed radiation points, which can provide flexible LoS links and diverse target observations \cite{li2026isac}. However, a favorable PA position does not always produce a strong or distinctive sensing response. If the PA radiates in an unsuitable direction, the target echo may be weak and nearby targets may be harder to separate. PA geometry design adds pattern diversity to position diversity by creating different angular responses. By sequentially activating PAs with different designed patterns, PASS can collect complementary target echoes for sensing, localization, and tracking. Complementary angular responses can improve target detectability and help distinguish nearby targets.

PASS-assisted ISAC benefits from distributed radiation points, which can provide flexible LoS links and diverse target observations \cite{li2026isac}. As indicated by the geometry-aware channel model in \eqref{pinch_channel.eq}, the PA-target link depends not only on the target distance $D$, but also on its angular direction $(\theta,\phi)$ through $G_{\mathcal{G}}(\theta,\phi)$. For a known PA position and state, $D$, $\theta$, and $\phi$ jointly characterize the target location relative to the PA. Consequently, targets at similar distances but in different directions can exhibit distinguishable angular gain signatures. By switching among PA geometries, rotations, or orientations, PASS can generate complementary angular responses, thereby improving target illumination, localization and tracking.

\subsection{Geometry-Assisted Secure Transmission}

Secure transmission is another promising application of PA geometry design. Existing PASS designs support physical-layer security by improving the channel advantage of legitimate users and covert communication by reducing signal observability at unintended receivers, through PA position optimization and transmit power control \cite{zhu2026secure,jiang2026covert}. However, without proper geometry design, even a well-placed PA may leak energy toward an eavesdropper or warden. Geometry design can reduce this risk by focusing radiation on the legitimate user and limiting leakage to unintended areas. Moreover, it can strengthen the legitimate link, enabling secure transmission with less exposed radiation.

\subsection{Pattern-Aided NOMA Transmission}

% Non-orthogonal multiple access (NOMA)-assisted PASS has been studied by optimizing PA activation, PA placement, and power allocation to improve multi-user transmission \cite{wang2025antenna}. These designs use PA placement to create distinguishable user links, which helps determine user pairing, power allocation, and successive interference cancellation (SIC) decoding. However, when users are at similar distances from the waveguide but in different directions, position control alone may not provide a sufficiently clear channel difference. PA geometry design can add an angular dimension to NOMA transmission. A directional PA can strengthen the link toward one user while reducing radiation toward another sector, helping to create a more favorable channel ordering for user pairing and SIC. Therefore, geometry-aware radiation patterns can complement power allocation and PA placement in future NOMA-assisted PASS. Beyond NOMA, geometry-aware PAs can also support service-dependent unicast and multicast transmission. For example, a near-isotropic circular PA can provide broad angular coverage for a multicast user group, whereas a more directional PA can concentrate radiated energy toward an intended user for unicast transmission.

Non-orthogonal multiple access (NOMA)-assisted PASS has been studied by jointly optimizing PA activation, PA placement, and power allocation for multi-user transmission \cite{wang2025antenna}. Existing designs exploit PA placement to create distinguishable user links, thereby facilitating user pairing, power allocation, and successive interference cancellation (SIC) decoding order. However, when users have similar PA-user distances but lie in different angular directions, PA placement alone may not create sufficient channel disparity. PA geometry design introduces an additional angular degree of freedom for NOMA transmission. A directional PA can enhance the link toward one user while suppressing radiation toward another sector, thereby creating a more favorable channel ordering for user pairing and SIC. Therefore, geometry-aware radiation patterns can complement PA placement and power allocation in future NOMA-assisted PASS. The same geometry-aware design principle can also support service-dependent unicast and multicast transmission. For example, a circular PA with a broader radiation pattern within the considered plane is suitable for providing broad angular coverage to a multicast user group, whereas a more directional PA can concentrate radiated energy toward an intended unicast user.

\section{Conclusion and Open Challenges}
In this article, we investigated geometry-dependent radiation in PASS. Starting from the physical principle and channel model, full-wave simulations and a 60 GHz prototype demo are presented to demonstrate that PA geometry and orientation can reshape radiation patterns and affect received signals. These findings indicate that PA geometry should be considered alongside PA location as a controllable design dimension, with the discussed application scenarios highlighting its potential for future PASS deployments. Since PASS is still at an early stage of development, substantial research challenges remain to be addressed before its full potential can be realized. The following key challenges and promising research directions are identified.
\begin{itemize}
        \item \textbf{Geometry-Aware Channel Modeling and Validation:} A key challenge is to develop tractable geometry-aware channel models. While full-wave simulations can reveal the radiation behavior of a specific PA geometry, system-level optimization requires compact models that capture the effects of PA shape, orientation and coupling strength. Measurements are also needed to validate these models in realistic wireless communication scenarios.
        \item \textbf{ Geometry-Control Hardware Co-Design:} Another important challenge lies in geometry-control hardware design. Future directional PASS prototypes require controllable structures for PA activation, rotation, and orientation adjustment. However, supporting materials and actuators placed near the waveguide may disturb the evanescent field and change the coupling behavior. Therefore, PA geometry, dielectric materials, and control structures should be jointly designed to ensure reliable reconfiguration without degrading the desired radiation pattern.
        \item \textbf{Beam Training and Channel Estimation for Directional PAs:} Geometry-aware directional PAs also bring new challenges to channel estimation and beam training. Existing PASS beam training mainly searches over PA locations or activation states. With directional PAs, the system must further identify suitable radiation patterns, rotation angles, and installation orientations. Compact pattern codebooks and efficient training are needed to exploit directional gain with limited pilot overhead.
\end{itemize}

% \section*{Acknowledgments}
% This should be a simple paragraph before the References to thank those individuals and institutions who have supported your work on this article.

%{\appendices
%\section*{Proof of the First Zonklar Equation}
%Appendix one text goes here.
% You can choose not to have a title for an appendix if you want by leaving the argument blank
%\section*{Proof of the Second Zonklar Equation}
%Appendix two text goes here.}

\bibliographystyle{IEEEtran}
\bibliography{refPAmagazine}

\end{document}